**WRF Simulation, Model Sensitivity, and Analysis of the**

**December 2013 New England Ice Storm**

Julia M. Simonson, Sean D. Birkel, Kirk A. Maasch, Paul A. Mayewski, Bradfield Lyon

Climate Change Institute, and School of Earth and Climate Sciences, University of Maine, Orono, Maine

Andrew M. Carleton

Department of Geography, and The Polar Center, Pennsylvania State University, University Park, Pennsylvania

Corresponding author: Julia Simonson, julia.simonson@maine.edu

ABSTRACT

Ice storms pose significant damage risk to electric utility infrastructure. In an attempt to improve storm response and minimize costs, energy companies have supported the development of ice accretion forecasting techniques utilizing meteorological output from numerical weather prediction (NWP) models. The majority of scientific literature in this area focuses on the application of NWP models, such as the Weather Research and Forecasting (WRF) model, to ice storm case studies, but such analyses tend to provide little verification of output fidelity prior to use. This study evaluates the performance of WRF in depicting the 21-23 December 2013 New England ice storm at the surface and in vertical profile. A series of sensitivity tests are run using eight planetary boundary layer (PBL) physics parameterizations, three reanalysis datasets, two vertical level configurations, and with and without grid nudging. Simulated values of precipitation, temperature, wind speed, and wind direction are validated against surface and radiosonde observations at several station locations across northeastern U.S. and southeastern Canada. The results show that, while the spatially and temporally averaged statistics for near-surface variables are consistent with those of select ice-storm case studies, near-surface variables are highly sensitive to model when examined at the station level. No single model configuration produces the most robust solution for all variables or station locations, although one scheme generally yields model output with the least realism. In all, we find that careful model sensitivity testing and extensive validation are necessary components for minimizing model-based biases in simulations of ice storms.

## 1. Introduction

While harsh winters are common in northern New England, damaging ice storms that impart significant cost to civil infrastructure and the regional economy are relatively rare. The most impactful ice storm in the region in recent history occurred 5-10 January 1998, resulting in over $1.4 billion in damage in the U.S. and southeastern Canada (Lott and Ross 2006). Another significant ice storm swept across the region on 21-23 December 2013. This more recent ice event was less severe than its 1998 counterpart, but nonetheless imparted costly damage to the regional electric grid: storm damage exceeded $1.9 million in Maine (Brogan 2014) and nearly $6.5 million in Vermont (NCEI 2014). The potential for extensive infrastructure damage, and uncertainty related to how climate change will affect the frequency and intensity of ice storms, warrants close inspection of how well numerical forecast models are able to depict and predict these events.

Most ice storm case studies focus primarily on the development of ice accretion modeling and forecasting methods. In these existing studies, forecast or reanalysis output is downscaled using a numerical weather prediction model (NWP), such as the Weather Research and Forecasting (WRF) model (Skamarock et al. 2008), which provides meteorological input for an ice accretion model. This approach has been utilized for predicting ice accretion on power lines (DeGaetano et al. 2008; Arnold 2009; Musilek et al. 2009; Pytlak et al. 2010; Hosek et al. 2011; Pytlak 2012; Zarnani et al. 2012), as well as in-cloud icing on wind turbines (Davis et al. 2013) and on other ground based structures in mountainous terrain (Nygaard et al. 2011). Outside of energy production and distribution industries, WRF has also been utilized to examine the role of sea surface temperatures in the Gulf of Mexico on ice-storm severity in the U.S. southern Great Plains (Mullens et al. 2016).

Despite the ubiquitous application of WRF for developing and validating ice forecasting systems, relatively few ice storm case studies are found in the literature compared to other modeled weather events. Those that are available concentrate on the ice forecast component, with limited consideration for the realism of the driving atmospheric model. Documentation of WRF output validation for ice storm case studies, if included at all, is generally restricted to spatially and temporally averaged statistical analyses of surface variables, as in Musilek et al. (2009) and Pytlak et al. (2010). Sensitivity tests are not typically reported, except with regards to ice-accretion modeling applications for a select number of physics parameterizations. For example, Nygaard et al. (2011) compared the performance of three cloud microphysics parameterization schemes for predicting supercooled cloud liquid water content and diagnosing median volume droplet diameter, two necessary input variables for ice accretion models. Eight WRF simulations centered on Mount Ylläs in northern Finland were evaluated using twice daily soundings from a meteorological observatory located 100 km east of the mountain. Modeled sounding profiles were considered representative of the atmospheric conditions during the simulations, with a simulation mean average error of 1.6°C. However, discrepancies arose when WRF was unable to resolve strong surface-based temperature inversions, resulting in 5°C overestimated modeled surface temperatures. Davis et al. (2013) also produced an icing study with sensitivity tests, wherein WRF was used to provide meteorological conditions at a Swedish wind farm for a wind turbine ice accumulation model. The sensitivity tests included three planetary boundary layer (PBL) schemes and three cloud microphysics schemes. General model performance was validated against 2-meter temperature and 10-meter wind speed observations from three surface stations, as well as temperature and wind speed observations at 80 meters at the wind farm. It was noted that while observed and modeled temperatures at the wind farm were in good agreement for temperatures above 0.5°C, large cold biases occurred when simulated temperatures were below freezing. The largest deviations in temperature occurred below -10°C, although this would not have a large impact on the ice model due to the particles freezing before contact with the turbine blades. These two studies provide more robust descriptions of model performance compared to other ice model studies, but nevertheless the evaluation of WRF output is brief and secondary to the desired model application.

This lack of model validation is in stark contrast to operational weather forecasting centered icing studies. Ikeda et al. (2013, 2017) used surface and sounding observations to assess the High-Resolution Rapid Refresh (HRRR) model in identifying surface precipitation phase for several case study ice storms effecting the central and eastern US. The HRRR is an operational NWP model that is built upon the WRF model and includes a postprocessing routine that identifies the type of precipitation at the surface. The authors found that the size and organization of weather systems is a factor in the forecast skill for precipitation extent and phase, with greater skill for larger, more organized systems compared to smaller events. For most events, the simulated near-surface temperatures had biases of less than 2°C, while several smaller events associated with cold-air damming on the eastern size of the Appalachian Mountains either did not have a subfreezing surface layer, or exhibited significant warm biases of up to 4°C within the layer. Overall, the study found that simulated locations and spatial extents of freezing rain were reasonable, but not nearly as robust as simulated depictions of rain and snow.

The sparse documentation on relevant WRF performance and sensitivity presents challenges for those interested in simulating ice storms with the greatest accuracy possible. Conditions are conducive for freezing rain when the atmosphere is highly stratified: a warm (above freezing) and moisture-laden air mass overruns a colder, subfreezing surface layer of air. Previous studies have determined that precipitation type is largely dependent on the maximum temperature of the warm layer, which is proportional to layer depth (Stewart and King 1987; Zerr 1997). Warm layers with maximum temperatures > 3°C allow for complete melting of snowflakes that fall through the layer, while lower temperatures allow partial or very little melting. The depth of the cold layer, which usually only extends 300-1200 m above the surface (Young 1978), is also crucial. Underestimating the depth of the cold layer would result in rain that would not freeze on contact, while overestimating the depth could result in the identification of sleet or ice pellets, as the rain refreezes before reaching the ground (Forbes et al. 1987). Considering that changes in temperature as low as 0.5°C can alter precipitation type (Thériault et al. 2010; Reeves et al. 2014), sensitivity testing is a crucial first step when using NWP models for research and development purposes in order to minimize the contribution of model-based uncertainty to icing forecasts.

This paper reports a suite of WRF sensitivity experiments designed to investigate the variability of model output to model configuration for the specific case of the December 2013 New England ice storm. The experiments test the impact of a variety of configuration options including the choice of PBL physics parameterization, reanalysis forcing, use of grid nudging, and the number of vertical levels. For lateral boundary forcing we utilize the reanalysis models ECMWF ERA-Interim (ERAI), ECMWF ERA5, and the NCEP North American Regional Reanalysis (NARR). Because ice accretion models utilize simulated values of air temperature, precipitation rate, and wind speed, we validate these simulated variables against surface and radiosonde observations. The tests reported here provide further insight into the sensitivity of WRF output to changes in model setup, thus providing general guidance for future WRF-based numerical simulations of ice storms.

This paper is structured as follows. The December 2013 New England ice storm is summarized in Section 2, with the data and model setup used described in Section 3. The results are described and discussed in sections 4 and 5. A summary of our major conclusions is presented in section 6.

## 2. December 2013 New England Ice Storm Case Study

The December 2013 New England ice storm was part of a larger storm system that brought freezing rain and heavy snow to the Midwest and Northeast, and tornadoes in the Southeast U.S. from 19 December through 23 December (NCDC 2014). This storm exhibited many of the same large-scale features present in the 1998 ice storm, as detailed by Gyakum and Roebber (2001) and Roebber and Gyakum (2003): a cold anticyclone in Canada, an anticyclone in the southwestern North Atlantic, and an inverted trough stretching from the Gulf of Mexico towards the Great Lakes (Fig. 1). A quasai-stationary front extended from east Texas through the Ohio Valley into New England, parallel to the southwesterly flow aloft. The air mass ahead of the front was unseasonably warm and moist for the time of year, with precipitable water values greater than 30 mm and a temperature gradient of more than 25°F (14°C) across the front. Strong low-level convergence and frontogenetic forcing ahead of a surface low resulted in heavy rain and tornadoes ahead of the front during 21 and 22 December, while behind the front fell heavy snow and freezing rain. Ice storm warning criteria (> 0.25 in ice accumulation) were met for counties in Texas, Oklahoma, Kansas, Iowa, Illinois, Michigan, New York, Vermont, and Maine during the storm.

In northern New York and New England, precipitation developed in two separate waves, with the location of the quasi-stationary front a key factor in the type of precipitation (Fig. 2). The first wave of precipitation lasted from approximately 1200 UTC 21 December until 1800 UTC 22 December. At this time, the front was largely stationary over northern New York, Vermont, and New Hampshire through the southeastern (or “Downeast”) coast of Maine, running parallel to the southwestern flow aloft. The front was partially obscured due to topographical features, with southwesterly flow over the White Mountains of New Hampshire and northeasterly flow to the east of the Longfellow Mountains in Maine. Several weak areas of low pressure tracked along the stationary front, with precipitation falling as rain over the Adirondacks and the White Mountains, freezing rain to the southeastern St Lawrence Valley and the northern Champlain Valley as well as Downeast Maine, mixing with sleet into central Maine and transitioning to snow to the north. The heaviest precipitation accumulations occurred during the latter half of this period, coinciding with a period of strong frontogenesis aloft ahead of the approach of a stronger low pressure system (Fig. 3a). The second wave of precipitation lasted from approximately 0000 UTC 23 December to 0000 UTC 24 December, during which the front drifted southward over eastern Massachusetts and Rhode Island and the low tracked across the Gulf of Maine. High temperatures on the 23$^{rd}$ range from -9°C (16°F) along the U.S-Canadian border to near 20°C (68°F) in parts of southern New England. This system brought rainfall to southern New England and additional freezing rain to the Downeast coast associated with moderate frontogenetic forcing aloft (Fig. 3b). Storm total ice accumulations as high as 32 mm (1.25 in) were reported in New York and Vermont and 25 mm (1.0 in) in Maine (Fig. 4). More than 75,000 customers in Vermont and 170,000 in Maine, as well as 66,000 in New York, lost electric service as a result of wire icing and downed trees, in some places for more than a week (NCDC 2014; NCEI 2014). Recovery efforts were hampered by extended extreme cold conditions and subsequent winter storms in the weeks following the ice storm, resulting in additional power outages.

## 3. Data and Model Setup

Simulations of the December 2013 New England ice storm were conducted using WRF version 3.9. Two one-way nested domains were used with grid spacings of 9 km and 3 km (Fig. 5). The simulations were initialized at 0000 UTC 20 December 2013 and ended at 0000 UTC 25 December 2013, with the first 24 hours used for model spinup. The model top was set to 50 hPa. Base physics options used for all sensitivity tests included the WRF single-moment 6-class microphysics scheme (Hong and Lim 2006), the Rapid Radiative Transfer Model for general circulation models (RRTMG) longwave radiation scheme (Iacono et al. 2008), the Goddard shortwave radiation scheme (Chou and Suarez 1999), the Kain-Fritsch cumulus scheme (Kain 2004) for the outer domain, and the Noah land surface model (Tewari et al. 2004). Preliminary simulations were run on the NCAR Yellowstone (Computational and Information Systems Laboratory 2016) supercomputer using 64 cores prior to its decommissioning. All sensitivity simulations reported here were conducted on the NCAR Cheyenne (Computational and Information Systems Laboratory 2017) supercomputer using 72 cores.

The model sensitivity tests consist of two groups with the configurations listed in Table 1. The first experiment group tests the WRF model sensitivity to choice of PBL scheme and the respective surface layer. We tested eight WRF PBL schemes, of which five of the eight PBL schemes utilize the Eta (Janjić 2002) and Revised MM5 (Jiménez et al. 2011) surface layer schemes, while the remaining PBL schemes were paired with their respective surface layer scheme. The two main components in which the schemes differ are in the order of closure and the extent of vertical mixing. The YSU and ACM2 schemes are first order closure schemes, in which higher order terms in the decomposed equations of motion are represented in terms of the mean. The remaining PBL schemes are 1.5 order, which predict higher order variables such as turbulent kinetic energy (TKE) by diagnosing second order (variance) moments for specific variables. Local mixing schemes allow only adjacent levels to influence variables at a given location, while non-local schemes include multiple levels. Most of the tested PBL schemes use local mixing, with two hybrid schemes (ACM2 and TEMF) utilizing either non-local or local mixing depending on the atmospheric stability, and YSU as the sole nonlocal scheme. PBL schemes also differ in relation to specific formulations, such as the incorporation of countergradient correction terms. More detailed descriptions of the PBL schemes tested in this study are found in Cohen et al. (2015) and Banks et al. (2016). For these simulations, initial and boundary conditions were supplied by ERA-Interim reanalysis (ERAI; ECMWF 2009), grid nudging (Stauffer and Seaman 1994) was applied to all levels for the outer domain with the nudging coefficients set to $3 \times 10^{-4}$, and 36 vertical levels were utilized.

The second experiment group tests the WRF model sensitivity to choice of several setup options using the MYJ PBL simulation as the "control". Three simulations test the use of grid nudging and the number of vertical levels. The lowest eta levels for the simulations using 36 vertical levels are 1.0, 0.993, 0.983, 0.97, 0.954, 0.934, 0.909, 0.880, 0.842, and 0.804, and the lowest eta levels for the simulations using 46 vertical levels are 1.0, 0.998, 0.995, 0.993, 0.988, 0.984, 0.98, 0.975, 0.97, 0.962, 0.954, 0.944, 0.934, 0.922, 0.909, 0.895, 0.88, 0.861, 0.842, and 0.804. Four simulations were run using the North American Regional Reanalysis (NARR; NCEP 2005) and ERA5 (ECMWF 2017) datasets, both with and without grid nudging. Compared to ERAI, a global reanalysis dataset available at 6 hourly intervals with 80 km grid spacing and 60 vertical levels, NARR has both a higher horizontal and temporal resolution (32 km and 3 hour, respectively), but fewer model levels (45). ERA5, the successor of ERAI, is a fifth-generation reanalysis produced by the ECMWF, with 31 km grid spacing, 137 vertical levels, and hourly output fields.

WRF model output was validated against surface station observations and tropospheric sounding data over 21-23 December 2013, when conditions were conductive for freezing rain. Hourly surface observations from 20 Automated Surface Observing System (ASOS) sites were obtained from the Iowa Environmental Mesonet website (https://mesonet.agron.iastate.edu/request/download.phtml), and sounding data were obtained from the NOAA/ESRL Radiosonde Database (https://ruc.noaa.gov/raobs/) for 6 sites (Fig. 5). Surface station and sounding sites within and without the ice storm extent were chosen to compare PBL scheme and overall WRF performance for icing and non-icing conditions. The statistical analysis was generally modeled after Musilek et al. (2009) and Pytlak et al. (2010), which included domain-wide metrics of hourly 2-meter temperature and 10-meter wind speed, as well as 6-hour accumulated precipitation. Statistical metrics from these two studies were used on the innermost domain over 21-23 December and include the mean error (bias), mean absolute error (MAE), and linear correlation coefficient (R). Statistics were also calculated for hourly 10-meter wind direction, as well as values of temperature, wind speed and direction from soundings at 0000 UTC, 0600 UTC (when available), and 1200 UTC. For the sounding variables, the WRF values were interpolated to the mandatory and observed significant levels below 700 hPa. The associated equations are as follow:

$$Bias = \frac{1}{N}\sum_{i=1}^{N}\theta_i$$

$$MAE = \frac{1}{N}\sum_{i=1}^{N}|\theta_i|$$

$$R = \frac{\sum_{i=1}^{N}[(O_i - \bar{O})(M_i - \bar{M})]}{(N-1)(\sigma_O \sigma_M)}$$

where

$$\theta_i = M_i - O_i$$

represents the deviation between the modeled and observed values of a particular variable, $\theta$, with $\bar{M}$ and $\bar{O}$ representing the modeled and observed 3-day averages (respectively), and N is the number of model-observation value pairs. Because wind direction is a circular variable and the absolute deviation cannot exceed 180, the difference between the modeled and observed wind direction is given following Carvalho et al. (2012):

$$\theta_i = (M_i - O_i)\left(\frac{1-360}{|M_i - O_i|}\right), if\ |M_i - O_i| > 180°$$

A positive (negative) bias represents a clockwise (counter-clockwise) deviation in modeled wind direction compared to the observed values. Domain-wide statistics for each PBL scheme simulation include modeled and observed values from all surface or sounding stations. To determine whether the simulations are significantly different from one another, a two-tailed paired t test (Wilks 2011) was performed against every variable of each simulation within the two groups. The statistical metrics detailed above, as well as the figures in the following section, were produced using the NCAR Command Language (NCL) version 6.4.0 (NCAR 2017).

## 4. Results

### *a. Assessment of Large-Scale Features*

Before validating WRF performance compared to surface observations as in previous ice storm studies, it is crucial to first assess the ability of the model to replicate the large-scale conditions of the storm. This consists of two parts: evaluating whether the fields provided by the ERA Interim reanalysis (ERAI) – such as the large-scale circulation as well as the mid- and low-level temperature and moisture fields – then examining how WRF depicts the depth and intensity of the air masses across the front. If ERAI does not sufficiently replicate the broader conditions during the event, then the ability of WRF to resolve local-scale features has to be called into question. Furthermore, understanding how WRF depicts the movement of the warm and cold air masses will lend itself to assessing the model's sensitivity to configuration choices.

At the synoptic scale, ERAI is representative of the atmospheric circulation during the event, including reproducing key surface features such as the highs over Canada and the western Atlantic and the inverted trough. ERAI also reproduces the enhanced surface temperature gradient throughout the southeastern US and the location of the freezing line, as well as ample moisture ahead of the front consistent with the presence of two atmospheric rivers. The 850 and 925 hPa temperature fields are also consistent between upper air observations and the reanalysis, with the freezing line and the location of the front parallel to the prevailing southeasterly flow. Equivalent potential temperatures at 850 hPa, the approximate level of maximum temperatures in the warm air mass, of over 287 K (5°C) are present over New England and are reflected in the fields of the outer WRF domain (Fig. 6). This temperature configuration in the lower troposphere is characteristic of large scale ice storms, favoring the gradual transition from rain to freezing rain/sleet instead of a direct change over to snow. From this assessment, we conclude that the fields provided by ERAI sufficiently represent the synoptic-scale features of the storm.

In the following paragraphs we examine the WRF model representation of critical factors associated with the ice storm, including the depth and intensity of the warm and cold air masses. For this examination we utilize vertical cross sections, oriented roughly perpendicular to the movement of the front, and utilize several sounding stations for verification. Individual cross section plots in Figure 7 and Figure 8 are based on the MYJ PBL simulation. In further assessing WRF realism for the ice storm, the modeled 2-meter temperature fields are compared against fields from the Real-Time Mesoscale Analysis (RTMA, NOAA/NCEP 2019; Figure 10), a dataset used by forecasters at the National Weather Service for producing and verifying weather forecasts.

The December 2013 ice storm can be characterized as two separate episodes, however the time frame differs with that of the two precipitation waves as discussed at the beginning of Section 2. The first episode of the ice storm begins around 1800 UTC 20 December with the formation of a wedge of subfreezing air near the surface (Fig. 7a). The cold wedge advances southward as warmer air aloft is advected northward (Fig. 7b), then from 1200 UTC to 2100 UTC 21 December the cold air retreats northward (Fig. 7c). Although minimal precipitation is observed during this period, freezing rain was observed in central Maine. Station observations also note that mist and fog is present throughout Maine and into New Hampshire. The second episode begins as the cold wedge redevelops and quickly intensifies from 2100 UTC 21 December to 1400 UTC 22 December. This period is characterized by enhanced frontogenesis in advance of a low pressure system (Fig. 3a) and a steep frontal slope (Fig. 8a), followed by a re-invigorated overrunning above the cold air wedge (Fig. 8b). The highest hourly rate of precipitation accumulation in northern New England occurs during this time frame (Fig. 9a,b). The second wave of precipitation, as shown in Figure 9c and 9d, occurs as the subfreezing surface layer is thinned (Fig. 8c) and ends as surface winds shift to the northwest and temperatures drop below freezing. Reports of freezing rain during this period are concentrated over southern Maine and southeastern New Hampshire.

Overall, WRF is able to sufficiently depict the depth and intensity of the elevated warm layer for the MYJ PBL simulation. Modeled maximum temperatures reflect those of sounding observations, which are over 4°C in southeastern Maine during the waves of precipitation. This indicates that temperatures within the warm airmass are sufficient for falling hydrometeors in this layer to melt completely. However, temperature biases near the surface are evident as surface air masses transition in northern New England. Cold biases are prevalent at 0000 and 1200 UTC 21 December and at 1200 UTC 22, while a warm bias is present at 0000 UTC 22 December (Fig. 10). Modeled surface temperatures are more consistent with observations for 23 December, as temperature biases are less prevalent than the prior two days. The modeled profiles for the endpoints of the cross section at the Caribou and Brookhaven stations (Fig. 7 and Fig. 8) are representative of the observed conditions for the duration of the storm.

Based on the cross sections and near-surface temperature maps, the vertical temperature profile and tropospheric winds appear to be well represented overall by WRF. The modeled temperature profiles closely follow observations within the elevated warm layer, then model performance generally decreases downward, with the largest temperature departures at or just above the surface.

*b. Sensitivity to PBL Scheme*

In this section we assess the sensitivity of the model to the chosen PBL scheme. This is done through comparing domain-wide statistical analyses as used by previous ice storm modeling studies and investigating the spatial and temporal variability of PBL performance using surface time series, soundings, and cross sections.

The results of the 3-day domain-wide sensitivity analysis of modeled 2-meter temperature, 10-meter wind speed and direction, and 6-hour precipitation are shown in Table 2 and sounding error statistics for the surface to 700 hPa in Table 3. Overall, the bias metric indicates that the model tends to overestimate wind speed (at the surface and up to 700 hPa) and precipitation, while near-surface temperature was generally underestimated in five of the eight PBL simulations and overestimated above the surface for all. The variability in MAE values across PBL schemes is minimal, with the exception of temperature and precipitation from the TEMF scheme. Error values for modeled sounding temperatures are approximately 0.75°C lower compared to surface values and nearly half for wind direction. Wind speed errors are 1 $ms^{-1}$ greater from modeled sounding profiles than at the surface, although the increased magnitude of wind speeds above the surface largely accounts for the difference. Linear correlations between modeled and observed surface values are high for temperature and precipitation (0.8 to 0.9), and less for wind speed and direction (0.5 to 0.7). The r-values for sounding variables are greater for all three sounding variables compared to the corresponding surface variables, further indicating that the modeled conditions are more in line with lower tropospheric sounding observations than at the surface. From the paired t tests, all except one pair of simulations (BouLac-TEMF) for 2-meter temperature, all except two pairs (MYJ-TEMF and BouLac-TEMF) for 10-meter wind speed, 21 of the 28 pairs for 10-meter wind direction, and all pairs with the TEMF simulation for 6-hour precipitation are significantly different at the 1.0% level. For the sounding variables, all except two pairs (MYJ-QNSE and MYNN2-UW) for temperature, 22 of the 28 pairs for wind speed, and 4 pairs of simulations (YSU-MYJ, YSU-BouLac, MYJ-MYNN2, and MYNN2-BouLac) for wind direction are significantly different.

In comparing cross sections and surface time series for the PBL simulations, we find that the schemes generally exhibit the same systematic biases near the frontal boundary. Figure 11 shows time series of 2-meter temperature and 10-meter winds for Portland, ME; the closest surface station to the Gray sounding location. Overall, the MYJ, QNSE, and ACM2 simulations tend to result in lower surface temperatures than the other PBL schemes, with more pronounced cold biases and lesser warm biases. Similarly, the BouLac simulation tends to exaggerate warm biases. However, surface temperatures are over- or under-estimated in all of the simulations during the first episode. Where the PBL simulations differ is the timing and the magnitude of the warm bias in surface temperatures, which correspond to a simulated wind shift from north/northeast to south/southeast. The same behavior is present at the Augusta, Bangor, and Millinocket surface stations, although the temperature maximum at 0000 UTC 22 December is less pronounced for stations farther north of the front. The only simulation which differs significantly is the TEMF scheme, which exhibits an enhanced surface temperature gradient on 21 December similar in characteristic to the later episode (Fig. 12). Winds are substantially stronger within the subfreezing and above freezing air masses and the temperature gradient more pronounced compared to the other PBL simulations. These conditions persist into 22 December, resulting in overestimated temperatures by as much as 10°C up to 700 hPa at the Gray sounding site and enhanced precipitation accumulations compared to the other WRF simulations (Fig. 13).

As the temperature gradient strengthens on 22 December, the simulations can be sorted into two groups. The MYJ, QNSE, and ACM2 schemes tend to represent the front with a shallower slope and the southward advancement of the cold air mass at a uniform rate, while the other PBL simulations show the surface air mass stalling in southern New Hampshire before moving into southern New England. This delay results in extremely overestimated surface temperatures at the southern stations in which the front passes over (Fig. 14). To north of the front, the temperature time series for the PBL simulations follow observations but with a spread of 2-3°C between them.

*c. Sensitivity to Reanalysis and Model Setup*

This section examines the sensitivity of WRF to other factors besides physics options, in part to pinpoint the source of systematic biases. This group of sensitivity tests focus on several model setup options that were determined through preliminary simulations and kept constant throughout the PBL simulations. The tested setup options include choice of reanalysis, use of grid nudging versus no grid nudging, and the chosen number of vertical levels for the simulation. Three of the simulations are variations of the "control" setup using the ERAI dataset, the MYJ PBL scheme, 36 model levels, and grid nudging: one simulation using 36 model levels and no grid nudging, and two simulations using 46 model levels with and without grid nudging. Four simulations are driven by two other reanalysis datasets (NARR and ERA5), also with and without grid nudging.

Table 4 and Table 5 show the results of the 3-day domain-wide surface and sounding (surface to 700 hPa) error statistics, respectively, calculated for the model sensitivity simulations. As with the PBL simulations, MAE values are larger for near-surface temperature and wind direction and for wind speed above the surface. These values also vary within a similar range as the PBL error values: 1-2°C for temperature, 1-3 $ms^{-1}$ for wind speed, 20-30 degrees for wind direction, and 1-1.5 mm for precipitation. Although the NARR simulation without grid nudging shows higher biases and MAE values for temperature and wind, the values are not nearly as extreme as those for the TEMF PBL simulation. R-values are similarly higher for temperature and precipitation compared to wind speed and direction and are higher for sounding variables compared to those at the surface. The paired t tests indicate that all except one pair of simulations (ERA4 36-NARR 36) for 2-meter temperature, all except two pairs (ERAI 46N-NARR 36 and ERA4 36-NARR 36) for 10-meter wind speed, 21 of the 28 pairs for 10-meter wind direction, and 13 of the 28 pairs for 6-hour precipitation are significantly different at the 1.0% level. For the sounding variables, all except three pairs (ERAI 36N-ERAI 36, ERAI 36-ERAI 46N, and ERAI 46N-ERAI 46) for temperature, all except four pairs (ERAI 36/46 with ERA5 36 and NARR 36) for wind speed, and 8 pairs of simulations for wind direction are significantly different.

As with the PBL simulations, the model setup simulations show the same general behaviors, such as the systematic temperature biases during the first episode, but with varying magnitudes. NARR lateral boundary forcing tends to produce higher surface temperatures than ERAI for northern New England stations, whereas ERA5 tends to have slightly lower surface temperatures. All of the simulations overestimate near-surface temperatures between 1800 UTC 21 December and 0600 UTC December 22, and most tend to underestimate temperatures at 1200 UTC on 21 and 22 December (Fig. 15). However, the NARR simulations exhibit higher temperatures than any of the other simulations from 0000 UTC to 1200 UTC 21 December. This behavior is more pronounced at the Augusta and Bangor surface stations, and to a lesser extent farther north at Millinocket. Grid nudging has a clear impact on the timing and the magnitude of the warm bias for the ERAI and NARR simulations but not for those forced by ERA5. Also, the ERAI simulations with grid nudging exhibit the shallow frontal slope and southward movement of the cold air mass into southern New England as the QNSE and ACM2 schemes, while the simulations without nudging reflect the conditions as shown by the other PBL simulations. The two simulations with 46 vertical levels tend to have slightly higher surface temperatures (< 0.5°C) within the cold dome and slightly lower temperatures to the south and east but are otherwise identical to their 36 level counterpart simulations. All of the model setup simulations generally fall within the same temperature range to the north and south of the frontal boundary during the second episode.

Although the evolution of the ice storm is similar among the model setup simulations, there are notable differences between the simulations which can be traced back to the driving reanalysis dataset. For example, there is an area of persistently cold near-surface temperature anomalies over the St. Lawrence River delta in the simulations forced by the ERAI reanalysis (Fig. 16). These anomalies are due to the relatively coarse resolution of the reanalysis, in which a portion of the delta is classified as land while the higher resolution reanalyses categorize the region as water. As a result, surface temperatures are as much as 5°C lower within the river delta for WRF simulations forced with ERAI and the colder temperatures are advected southeastward into the river valley. Another example of notable temperature anomalies is between the NARR and ECMWF reanalysis simulations due to differences in land surface cover classification over lakes. Near-surface temperatures are often several degrees Celsius higher over open-water lakes than the surrounding area in the NARR simulations. These anomalies are especially prominent over Lake Champlain (Fig. 16a) and likely play a substantial role in the differing extent of sub-freezing temperatures within the valley between simulations.

*d. Model Sensitivity and Precipitation Type*

For this section, we examine how the variability of conditions between simulations can change the type of precipitation identified. Although the sensitivity simulations show variations in modeled temperature, wind speed and direction, and precipitation values, it is difficult to determine whether the variability would significantly alter precipitation type. WRF does not explicitly identify precipitation type, so previous modeling studies have required the use of outside classification algorithms. However, there are a wide variety of classification algorithms to choose from, each with their own advantages and disadvantages.

As the primary difference between the sensitivity simulations is the modeled temperature values, most prominently at the surface, we chose a simplified "top-down" approach based on the maximum temperature in the warm air mass, which is around 850 hPa, and the surface temperature. The precipitation categories and their respective temperature ranges are listed in Table 6. The categories are based on the assumption that cloud top temperatures are low enough (typically < -10°C) for the formation of ice crystals that fall into the warm layer, and that the air is saturated when precipitation occurs. Based on the observed and model soundings, as well as the IR images in Figure 1, we find this assumption to be valid for the case study ice storm. The maximum temperature of the warm layer then determines whether the ice crystals completely or partially melt, and the surface temperature determines if refreezing occurs before precipitation reaches the surface. While overly simplistic, this method provides a useful physical representation for each precipitation type: snow indicates the lack of a sufficient melting layer; mixed precipitation signifies partial melting in the warm layer or refreezing in the cold layer; freezing rain represents complete melting within the warm layer and the presence of a shallow subfreezing surface layer; and rain denotes the absence of a subfreezing surface layer.

The sensitivity of precipitation type to model setup depends on the timing of modeled temperature biases in relation to the precipitation field. The underestimated surface temperatures which peak at 1200 UTC 22 December result in minor differences in the transition from freezing rain to mixed precipitation with more noticeable cold biases (Fig. 17a,b). However, the impact to precipitation type for underestimated temperatures at 1200 UTC 21 December limited, as modeled precipitation is minimal in regions with high storm total ice accumulations. This is also true for 23 December, as the extent of freezing rain is more isolated than that of the first wave of precipitation (Fig. 17c,d). The overestimated modeled surface temperatures shown in all of the simulations around 0000 UTC 22 December have a more substantial effect, resulting in rain throughout Downeast Maine when observations report freezing rain (Fig. 17e,f). Simulations which exhibit a longer duration of above freezing surface temperatures would result in much lower ice accumulations during this period. Although the spatial extent of precipitation is generally consistent among most of the simulations, there are some distinct differences. For example, the ERA5 and ERAI simulations show a larger expanse of mixed precipitation within the Champlain Valley and in southern Maine while freezing rain is more widespread in the NARR simulations (Fig. 18). However, limited observations in this area preclude a more thorough assessment of precipitation classification accuracy.

## 5. Discussion

The sensitivity tests reported here indicate varying confidence in the fidelity of the WRF model solutions. WRF is generally reliable in reproducing the overall meteorological conditions associated with the December 2013 New England ice storm, where the model resolves most temperature inversions, as well as the large-scale movement of the storm system and the accompanying frontal boundary. However, near-surface temperatures close to 0°C are not sufficiently reproduced at the station scale to the precision required for accurate classification of precipitation type. Precipitation type algorithms, such as that used in the HRRR forecast model, and ice accretion models rely directly on modeled temperature, wind, and precipitation rate to produce icing forecasts. Variability between the simulations are relatively small in plan view, yet not so insignificant as to discount model setup as a factor in assessing forecast accuracy for ice storms. Although the WRF model itself does not explicitly identify precipitation type, we utilize a simplified precipitation classification approach to distinguish regions with a higher probability for freezing rain. Based on the surface temperature biases present, we postulate that the model output would slightly favor the misclassification of freezing rain as sleet or mixed precipitation for simulations that tend to underestimate surface temperatures on 22 and 23 December and that all simulations would favor rain over freezing rain around 0000 UTC 22 December.

It is difficult to identify one simulation that produces an overall robust solution for conditions both inside and outside of the ice storm. Although all of the sensitivity tests produce simulations that follow a similar progression of the ice storm, the model fields are not representative of observations throughout. Some of the simulations produce higher surface temperatures, which minimize the effect of cold biases during the weakly and strongly forced episode at the expense of producing a longer period of abnormally warm temperatures, favoring the identification of rain over freezing rain. Simulations with generally lower surface temperatures similarly favor an earlier transition to mixed precipitation. However, based on the statistical analysis alone, these tendencies are either obscured or less obvious. Our results do not indicate clear differences in the model solution solely by PBL scheme closure or vertical mixing Only the TEMF scheme stands out as a noticeable outlier, with especially high MAE values for precipitation and near-surface and lower tropospheric temperature. Further investigation was required to pinpoint the cause of the departures as enhanced frontogenesis on 21 December which can be attributed to the explicit inclusion of shallow cumulus within the scheme's formulation (Angevine et al. 2010). The NARR forced simulation without grid nudging was similarly an outlier for the model setup simulations, although the greater error values are attributed to systematically higher temperatures as opposed to a marked difference in the overall storm evolution. With the use of a single case study storm, we cannot determine whether the various model setups tested would perform similarly for another ice storm in this region. Our overall results do not afford a single "best" model configuration; instead, the combine results of the sensitivity tests reflect the interaction among various model components.

Although the scope of this study is limited to a single storm, the extensive validation and analysis of the December 2013 ice storm demonstrates the numerous challenges of modeling ice storms, from both a weather forecasting and research application perspective. Ikeda et al. (2013) note that while current NWP models are better able to predict the areal extent and timing of precipitation associated with large-scale cold season systems compared to warm season convective precipitation, even high resolution forecast models such as the HRRR have difficulty predicting the phase of precipitation for ice storms. As the classification of freezing rain and sleet is more sensitive to model uncertainty compared to rain and snow, the advantage of hindcast case studies is the ability to test a variety of model configurations to identify and minimize systematic model biases. However, previous modeling studies of ice storms using WRF rely on a single model setup and do not examine the ways in which their setup influenced the modeled meteorological conditions, and thus modeled ice accumulations. This and other case study simulations (e.g. Musilek et al. 2009; Pytlak et al. 2010) report similar mean errors of 1-2°C, and our results indicate several instances in which model setup choices can alter the type of precipitation identified from model output. Furthermore, the WRF results are corroborated solely by a set of surface observations sites, limiting the scope of WRF performance to point locations and not to the large or local-scale features of the individual case study storms. By not addressing the sensitivity of ice forecasting systems to the variable fields they are provided, these systems will require constant modification as NWP models are updated in order to compensate for a variety of uncertainties. These points could hamper the development of generally applicable ice accretion forecasting methods.

## 6. Conclusions

This study evaluates the sensitivity of the WRF mesoscale model to several model setup factors in a case study of the New England ice storm of 21-23 December 2013. Simulated values of 6-hour precipitation, as well as near-surface and vertical profiles of temperature, wind speed, and wind direction, are validated against surface station and radiosonde observations. Overall, we find that WRF produces robust depictions of key features of the ice storm, including the large-scale circulation and vertical structure of the atmosphere. The results of the simulations are also consistent with the results of previous ice storm case studies used to develop and run ice accretion models. However, near-surface temperatures vary at the local scale between the suite of sensitivity tests and are not obvious from the simulation average statistical analysis. We find that no single simulation produced high fidelity simulations of the ice storm overall, although the TEMF PBL scheme was clearly unsuitable for the examined weather event. This study underscores the importance of extensive validation and testing to assess the accuracy and realism of the WRF model solution in comparison to observational data, particularly for case studies of weather events as impactful to civil infrastructure as ice storms.

*Acknowledgments*. This study is supported by a University of Maine Signature and Emerging Area of Excellence Graduate Fellowship to the Climate Change Institute. High-performance computing support from Cheyenne and Yellowstone was provided by NCAR's Computational and Information Systems Laboratory, sponsored by the National Science Foundation.

Table 1. Summary of model simulations used in this study.

| Short Name | Reanalysis | PBL Scheme | Surface Layer | Nudging (Y/N) | Vertical Levels |
|---|---|---|---|---|---|
| YSU | ERAI | Yonsei University | Revised MM5 similarity | Y | 36 |
| ACM2 | ERAI | Asymmetric Convective Model Version 2 | Revised MM5 similarity | Y | 36 |
| MYJ | ERAI | Mellor-Yamada-Janjic | Eta similarity | Y | 36 |
| QNSE | ERAI | Quasi-Normal Scale Elimination | QNSE | Y | 36 |
| MYNN2 | ERAI | Mellor-Yamada-Nakanishi-Niino Level 2.5 | MYNN | Y | 36 |
| BouLac | ERAI | Bougeault-Lacarrere | Revised MM5 similarity | Y | 36 |
| UW | ERAI | University of Washington | Revised MM5 similarity | Y | 36 |
| TEMF | ERAI | Total Energy-Mass Flux | TEMF | Y | 36 |
| ERAI 36 | ERAI | Mellor-Yamada-Janjic | Eta similarity | N | 36 |
| ERAI 46N | ERAI | Mellor-Yamada-Janjic | Eta similarity | Y | 46 |
| ERAI 46 | ERAI | Mellor-Yamada-Janjic | Eta similarity | N | 46 |
| ERA5 36N | ERA5 | Mellor-Yamada-Janjic | Eta similarity | Y | 36 |
| ERA5 36 | ERA5 | Mellor-Yamada-Janjic | Eta similarity | N | 36 |
| NARR 36N | NARR | Mellor-Yamada-Janjic | Eta similarity | Y | 36 |
| NARR 36 | NARR | Mellor-Yamada-Janjic | Eta similarity | N | 36 |

Table 2. Statistical metrics of hourly 2-meter temperature (T2, N = 11,472), 10-meter wind speed (WS10, N = 11,416) and wind direction (WD10, N = 9,608), and 6-hour precipitation (PRE, N = 1,520) by PBL scheme simulation, averaged over all stations. Precipitation metrics include U.S. stations only.

| Variable | Scheme | Bias | MAE | R |
|---|---|---|---|---|
| T2 (°C) | YSU | -0.02 | 1.93 | 0.95 |
| | ACM2 | -0.52 | 1.85 | 0.96 |
| | MYJ | -1.18 | 1.89 | 0.97 |
| | QNSE | -1.10 | 1.87 | 0.97 |
| | MYNN2 | 0.33 | 2.06 | 0.95 |
| | BouLac | 1.00 | 2.12 | 0.95 |
| | UW | -0.29 | 2.15 | 0.95 |
| | TEMF | 1.14 | 3.14 | 0.92 |
| WS10 (m $s^{-1}$) | YSU | 0.49 | 1.70 | 0.57 |
| | ACM2 | 0.58 | 1.67 | 0.60 |
| | MYJ | 0.83 | 1.52 | 0.72 |
| | QNSE | 0.70 | 1.53 | 0.70 |
| | MYNN2 | 0.02 | 1.61 | 0.56 |
| | BouLac | 0.98 | 1.90 | 0.61 |
| | UW | 0.42 | 1.64 | 0.58 |
| | TEMF | 0.89 | 1.93 | 0.48 |
| WD10 (degrees) | YSU | 4.69 | 28.57 | 0.52 |
| | ACM2 | 5.00 | 24.25 | 0.51 |
| | MYJ | 4.05 | 22.42 | 0.57 |
| | QNSE | 0.46 | 23.48 | 0.55 |
| | MYNN2 | 5.48 | 28.96 | 0.48 |
| | BouLac | 6.83 | 31.20 | 0.46 |
| | UW | 2.38 | 26.28 | 0.47 |
| | TEMF | -0.29 | 28.48 | 0.48 |
| PRE (mm) | YSU | 0.50 | 1.00 | 0.80 |
| | ACM2 | 0.58 | 0.99 | 0.81 |
| | MYJ | 0.46 | 0.99 | 0.80 |
| | QNSE | 0.47 | 0.99 | 0.80 |
| | MYNN2 | 0.49 | 1.00 | 0.80 |
| | BouLac | 0.42 | 0.93 | 0.81 |
| | UW | 0.48 | 1.01 | 0.79 |
| | TEMF | 2.78 | 3.11 | 0.53 |

Table 3. Statistics of sounding temperature (T, N = 4,880), wind speed (WSP, N = 3,869), and wind direction (WDR, N = 3,869) for each PBL scheme simulation, averaged over all stations.

| Variable | Scheme | Bias | MAE | R |
|---|---|---|---|---|
| T (°C) | YSU | 0.40 | 1.24 | 0.98 |
| | ACM2 | 0.26 | 1.22 | 0.98 |
| | MYJ | 0.17 | 1.21 | 0.98 |
| | QNSE | 0.17 | 1.27 | 0.97 |
| | MYNN2 | 0.45 | 1.25 | 0.98 |
| | BouLac | 0.58 | 1.39 | 0.97 |
| | UW | 0.48 | 1.31 | 0.97 |
| | TEMF | 1.16 | 1.83 | 0.95 |
| WSP (m $s^{-1}$) | YSU | -0.15 | 2.86 | 0.91 |
| | ACM2 | 0.02 | 2.68 | 0.92 |
| | MYJ | 0.02 | 2.83 | 0.91 |
| | QNSE | 0.14 | 2.93 | 0.9 |
| | MYNN2 | 0.01 | 2.92 | 0.91 |
| | BouLac | -0.44 | 2.69 | 0.92 |
| | UW | 0.12 | 2.84 | 0.91 |
| | TEMF | 0.71 | 3.36 | 0.87 |
| WDR (degrees) | YSU | 0.93 | 10.41 | 0.66 |
| | ACM2 | 1.14 | 10.15 | 0.6 |
| | MYJ | 0.78 | 10.10 | 0.66 |
| | QNSE | 0.06 | 10.32 | 0.65 |
| | MYNN2 | 1.11 | 10.29 | 0.71 |
| | BouLac | 0.12 | 10.05 | 0.69 |
| | UW | 0.85 | 10.16 | 0.65 |
| | TEMF | -1.16 | 12.00 | 0.61 |

Table 4. As in Table 2, except by model setup simulation. The ERAI 36N simulation corresponds with the MYJ PBL simulation. Tests designated with an N denote grid nudging.

| Variable | Test | Bias | MAE | R |
|---|---|---|---|---|
| T2 (°C) | ERAI 36N | -1.18 | 1.89 | 0.97 |
| | ERAI 36 | -0.08 | 2.15 | 0.94 |
| | ERAI 46N | -0.98 | 1.81 | 0.97 |
| | ERAI 46 | 0.13 | 2.02 | 0.95 |
| | ERA5 36N | -1.13 | 1.88 | 0.96 |
| | ERA5 36 | -0.59 | 1.77 | 0.96 |
| | NARR 36N | 0.31 | 2.00 | 0.94 |
| | NARR 36 | 1.42 | 2.64 | 0.90 |
| WS10 (m $s^{-1}$) | ERAI 36N | 0.83 | 1.52 | 0.72 |
| | ERAI 36 | 1.72 | 2.33 | 0.53 |
| | ERAI 46N | 0.77 | 1.50 | 0.71 |
| | ERAI 46 | 1.60 | 2.25 | 0.53 |
| | ERA5 36N | 1.09 | 1.75 | 0.67 |
| | ERA5 36 | 1.36 | 1.94 | 0.68 |
| | NARR 36N | 0.35 | 1.59 | 0.63 |
| | NARR 36 | 2.00 | 2.71 | 0.44 |
| WD10 (degrees) | ERAI 36N | 4.05 | 22.42 | 0.57 |
| | ERAI 36 | 3.20 | 31.28 | 0.50 |
| | ERAI 46N | 2.32 | 22.70 | 0.56 |
| | ERAI 46 | 0.59 | 30.63 | 0.50 |
| | ERA5 36N | 0.79 | 24.77 | 0.52 |
| | ERA5 36 | 3.40 | 25.51 | 0.53 |
| | NARR 36N | -0.19 | 30.61 | 0.53 |
| | NARR 36 | 3.66 | 38.62 | 0.45 |
| PRE (mm) | ERAI 36N | 0.46 | 0.99 | 0.80 |
| | ERAI 36 | 0.84 | 1.51 | 0.61 |
| | ERAI 46N | 0.48 | 0.99 | 0.80 |
| | ERAI 46 | 1.03 | 1.63 | 0.58 |
| | ERA5 36N | 0.42 | 1.06 | 0.78 |
| | ERA5 36 | 0.93 | 1.28 | 0.80 |
| | NARR 36N | -0.29 | 1.03 | 0.59 |
| | NARR 36 | 0.49 | 1.04 | 0.81 |

Table 5. As in Table 3, except by model setup simulation.

| Variable | Test | Bias | MAE | R |
|---|---|---|---|---|
| T (°C) | ERAI 36N | 0.18 | 1.21 | 0.98 |
| | ERAI 36 | 0.17 | 1.43 | 0.97 |
| | ERAI 46N | 0.22 | 1.25 | 0.98 |
| | ERAI 46 | 0.32 | 1.56 | 0.96 |
| | ERA5 36N | -0.39 | 1.32 | 0.97 |
| | ERA5 36 | -0.21 | 1.35 | 0.97 |
| | NARR 36N | 0.50 | 1.40 | 0.97 |
| | NARR 36 | 0.80 | 1.90 | 0.94 |
| WSP (m $s^{-1}$) | ERAI 36N | 0.02 | 2.83 | 0.91 |
| | ERAI 36 | 0.92 | 3.29 | 0.89 |
| | ERAI 46N | 0.10 | 2.92 | 0.92 |
| | ERAI 46 | 1.22 | 3.53 | 0.89 |
| | ERA5 36N | -0.45 | 2.95 | 0.91 |
| | ERA5 36 | 0.83 | 3.08 | 0.90 |
| | NARR 36N | -1.48 | 3.46 | 0.88 |
| | NARR 36 | 1.38 | 3.63 | 0.86 |
| WDR (degrees) | ERAI 36N | 0.78 | 10.10 | 0.66 |
| | ERAI 36 | -1.32 | 11.02 | 0.62 |
| | ERAI 46N | 0.33 | 11.02 | 0.65 |
| | ERAI 46 | -0.31 | 12.80 | 0.59 |
| | ERA5 36N | -4.09 | 9.75 | 0.71 |
| | ERA5 36 | -2.74 | 11.15 | 0.65 |
| | NARR 36N | -1.53 | 12.56 | 0.64 |
| | NARR 36 | -0.84 | 14.17 | 0.61 |

Table 6. Temperature thresholds (850 hPa and surface) for precipitation classification. Mixed category includes frozen hydrometeors that result from partially melting (pm) in the warm layer and refreezing (rfrz) within the subfreezing surface layer. Threshold values based on Baumgardt (1999) and UCAR (2005).

| | T850 | Tsfc |
|---|---|---|
| Snow | < 1°C | < 0°C |
| Mix | 1°C – 3°C **(pm)**<br>or<br>> 3°C **(rfrz)** | < 0°C<br>or<br>< -6°C |
| Freezing Rain | > 3°C | -6°C – 0°C |
| Rain | > 3°C | > 0°C |

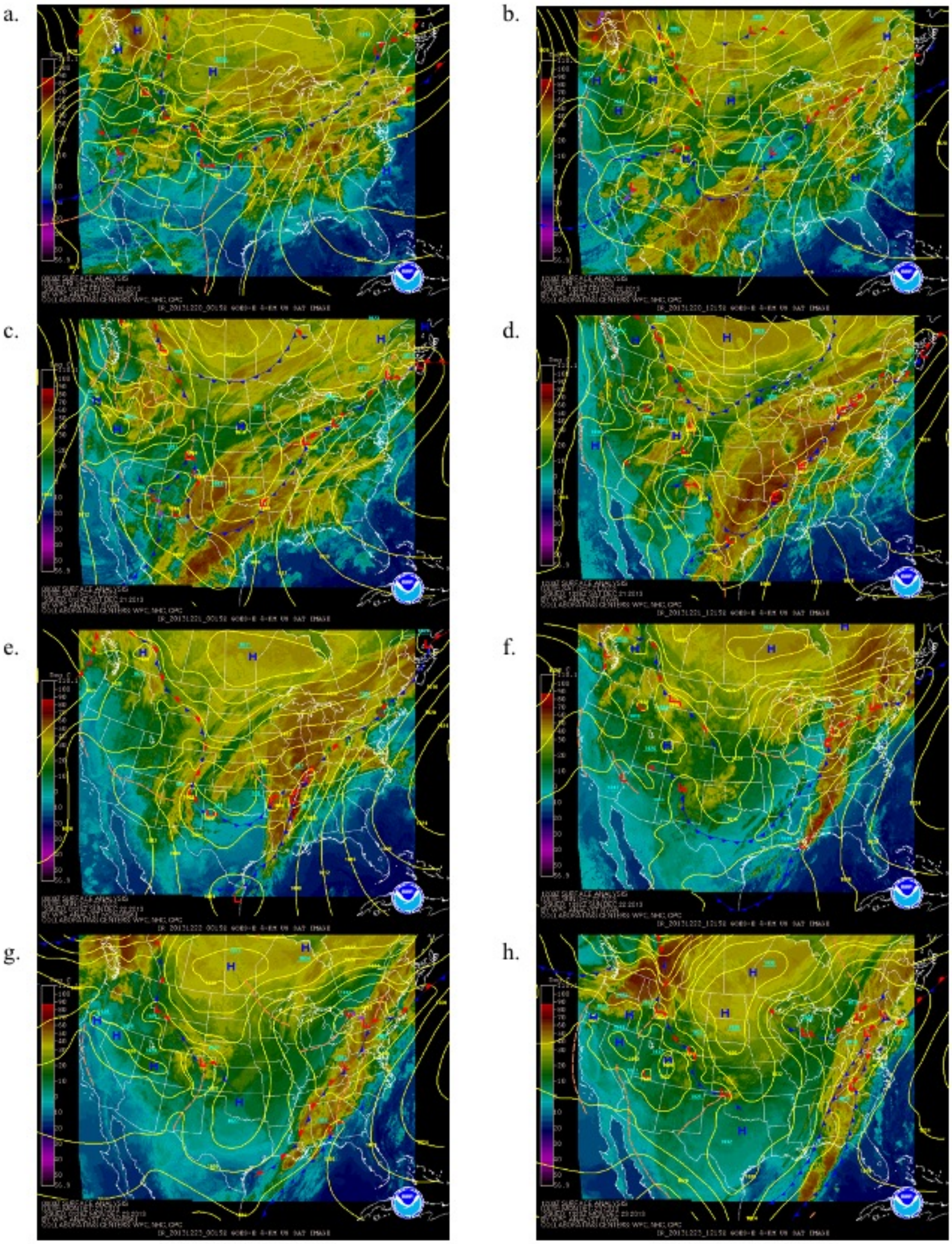

Fig. 1. Surface analysis/infrared satellite composites (Weather Prediction Center 2018) for 0000 UTC (a) 20, (c) 21, (e) 22, and (g) 23 Dec 2013 and 1200 UTC (b) 20, (d) 21, (f) 22, and (h) 23 Dec 2013.

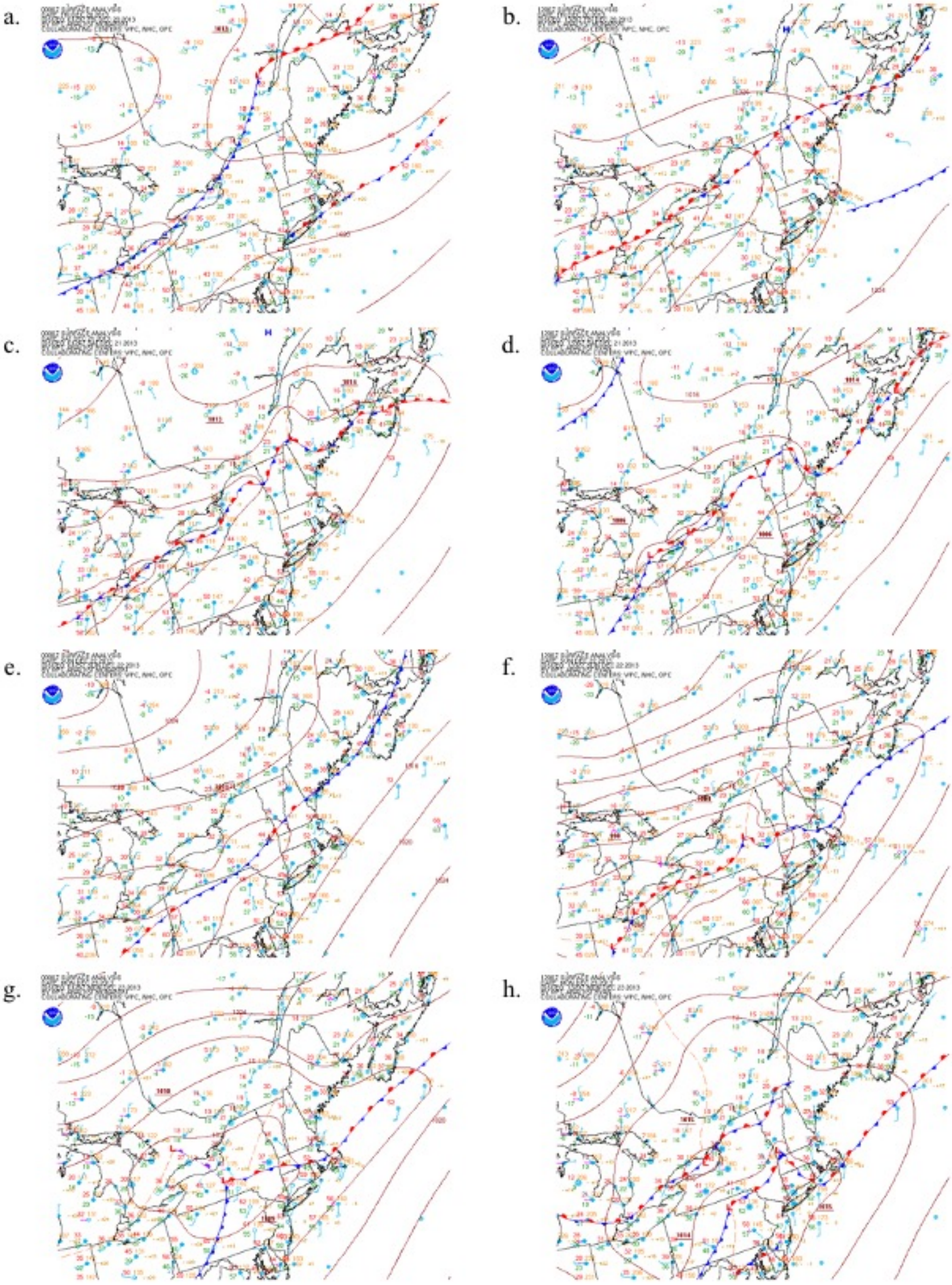

Fig. 2. Regional surface analyses same as Fig. 1.

a.

b.

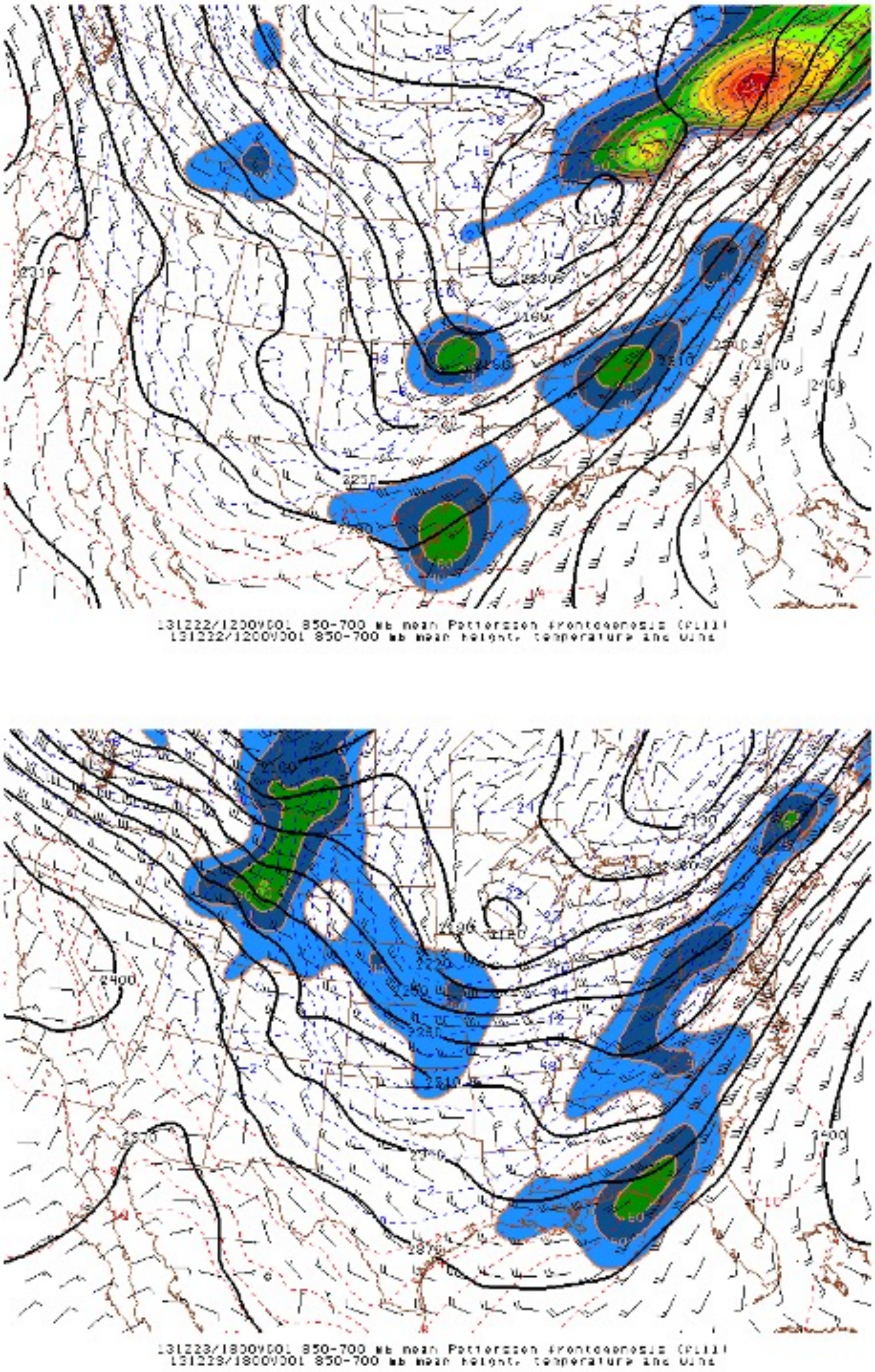

Fig. 3. Mesoscale analysis of 850 -700 hPa mean Petterssen frontogenesis, mean height, temperature, and wind for (a) 1200 UTC 22 Dec and (b) 1800 UTC 23 Dec from Storm Prediction Center (2019).

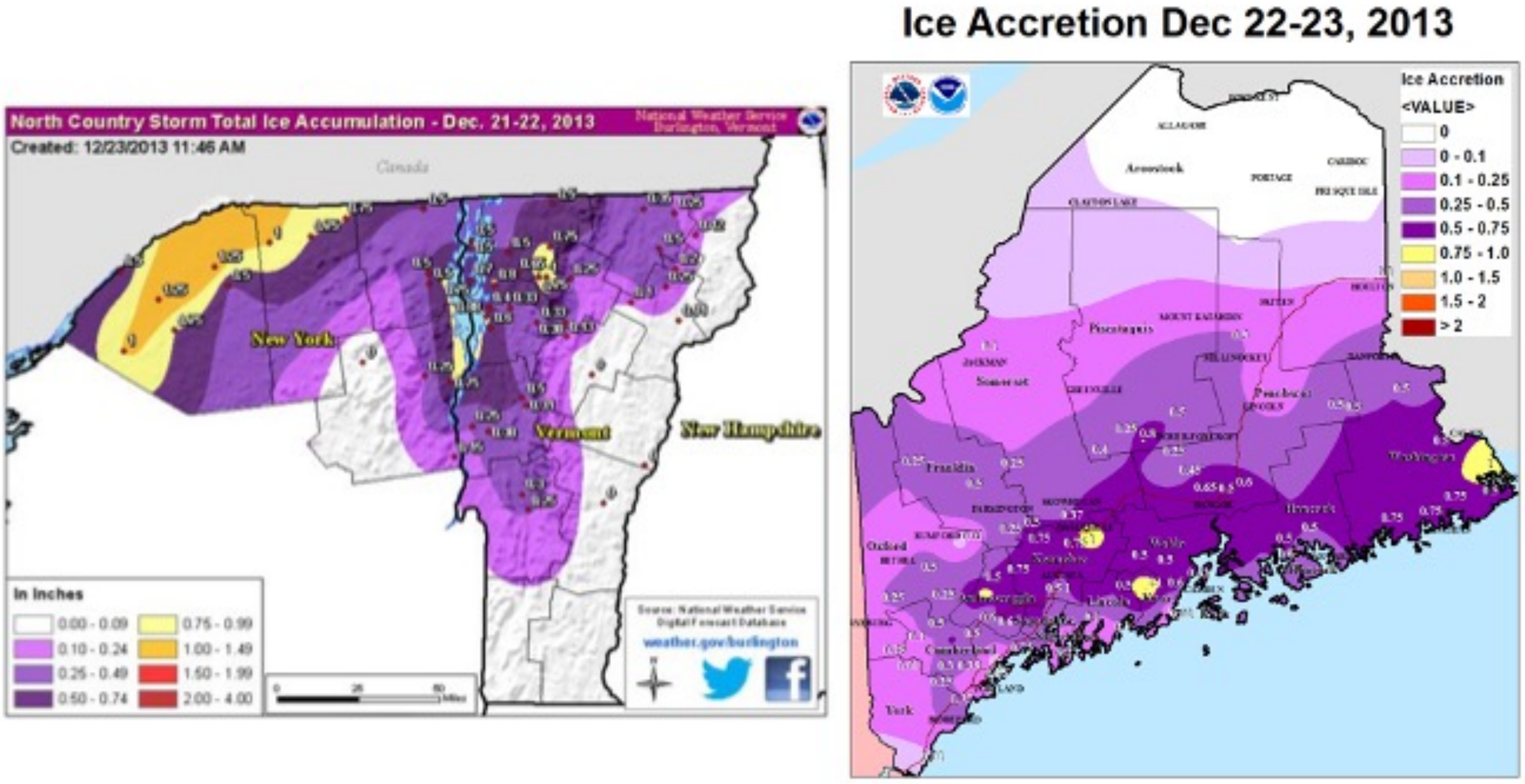

Fig. 4. Storm total ice accumulation maps [from (left) Taber (2015) and (right) NOAA (2019)].

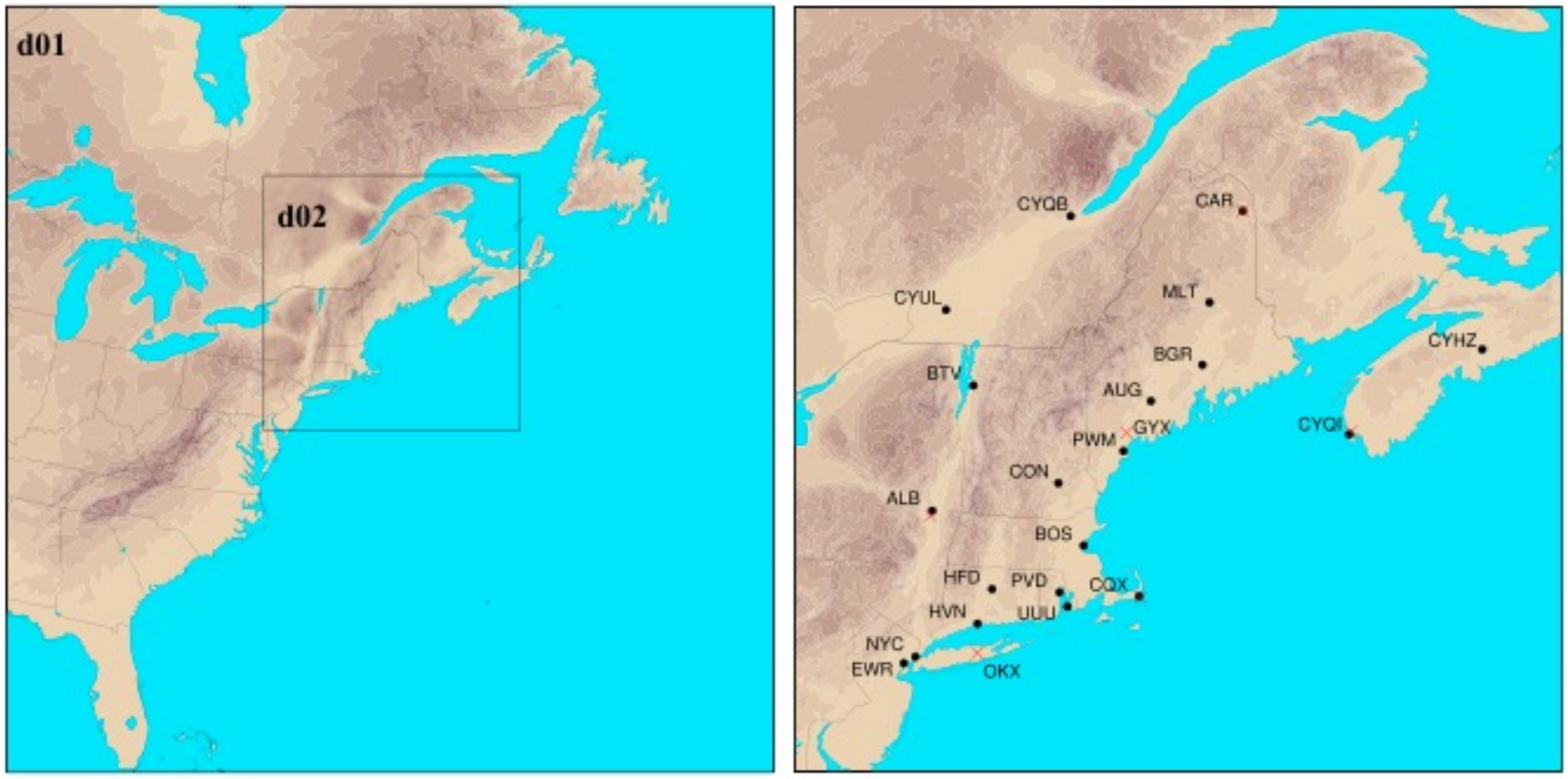

Fig. 5. WRF model domains (left) and locations of ASOS and radiosonde stations (right). Surface stations include Albany, NY (ALB), Augusta, ME (AUG), Bangor, ME (BGR), Boston, MA (BOS), Burlington, VT (BTV), Caribou, ME (CAR), Concord, NH (CON), Chatham, MA (CQX), Newark, NJ (EWR), Hartford, CT (HFD), New Haven, CT (HVN), Millinocket, ME (MLT), New York City, NY, (NYC), Providence, RI (PVD), Portland, ME (PWM), Newport, RI (UUU), Halifax, NS (CYHZ), Quebec City, QC (CYQB), Yarmouth, NS (CYQI), and Montreal, QC (CYUL). Sounding stations include Gray, ME (GYX) and Brookhaven, NY (OKX), as well as sites collocated with the Albany, Caribou, Chatham, and Yarmouth surface stations.

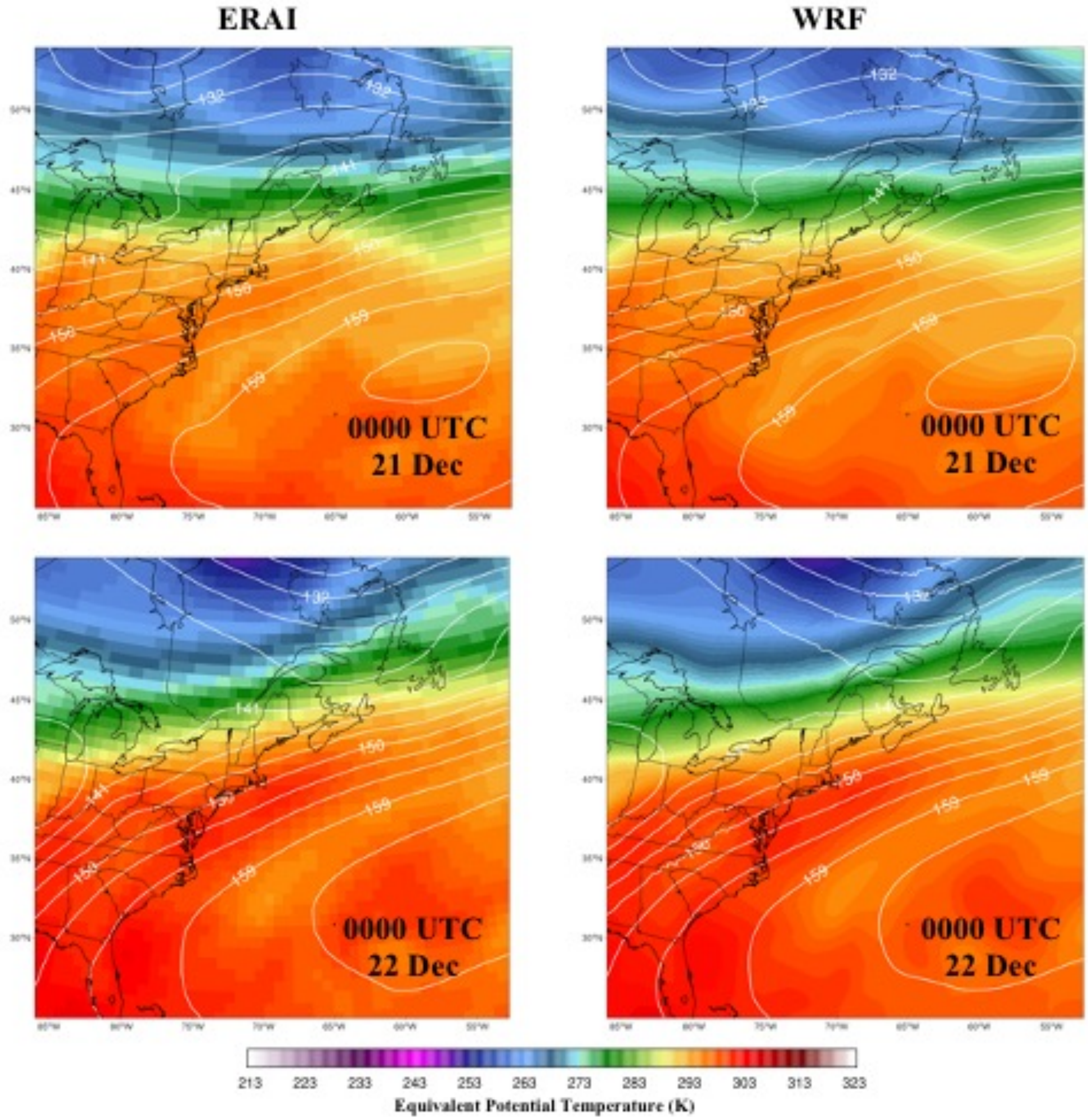

Fig. 6. Comparison of 850 hPa equivalent potential temperature (K) and geopotential height (dm) contours at 0000 UTC 21 Dec (top) and 0000 UTC 22 Dec (bottom) from ERAI (left) and the WRF outer domain for the MYJ PBL simulation (right).

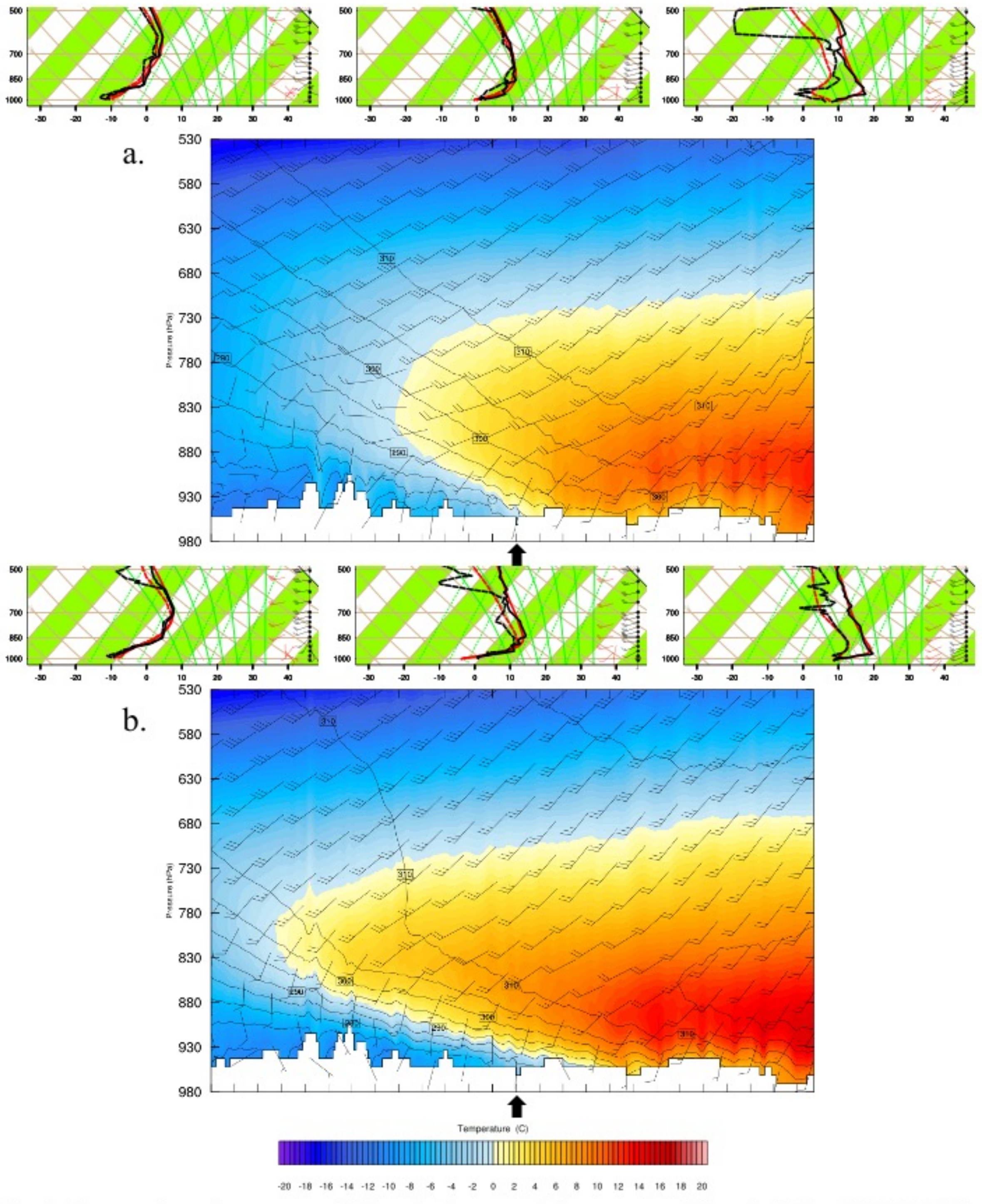

Fig. 7. Cross sections of temperature (°C), equivalent potential temperature (K), and winds at (a) 0000 UTC and (b) 1200 UTC 21 Dec, and (c) 0000 UTC 22 Dec for the MYJ PBL simulation. Observed (black) and WRF modeled (red) soundings for the Caribou (left), Gray (center), and the Brookhaven (right) stations are above each cross section, with the location of the Gray station designated by the black arrow. Temperature profile (solid) is plotted to the right of dewpoint profile (dashed), and winds are in $ms^{-1}$.

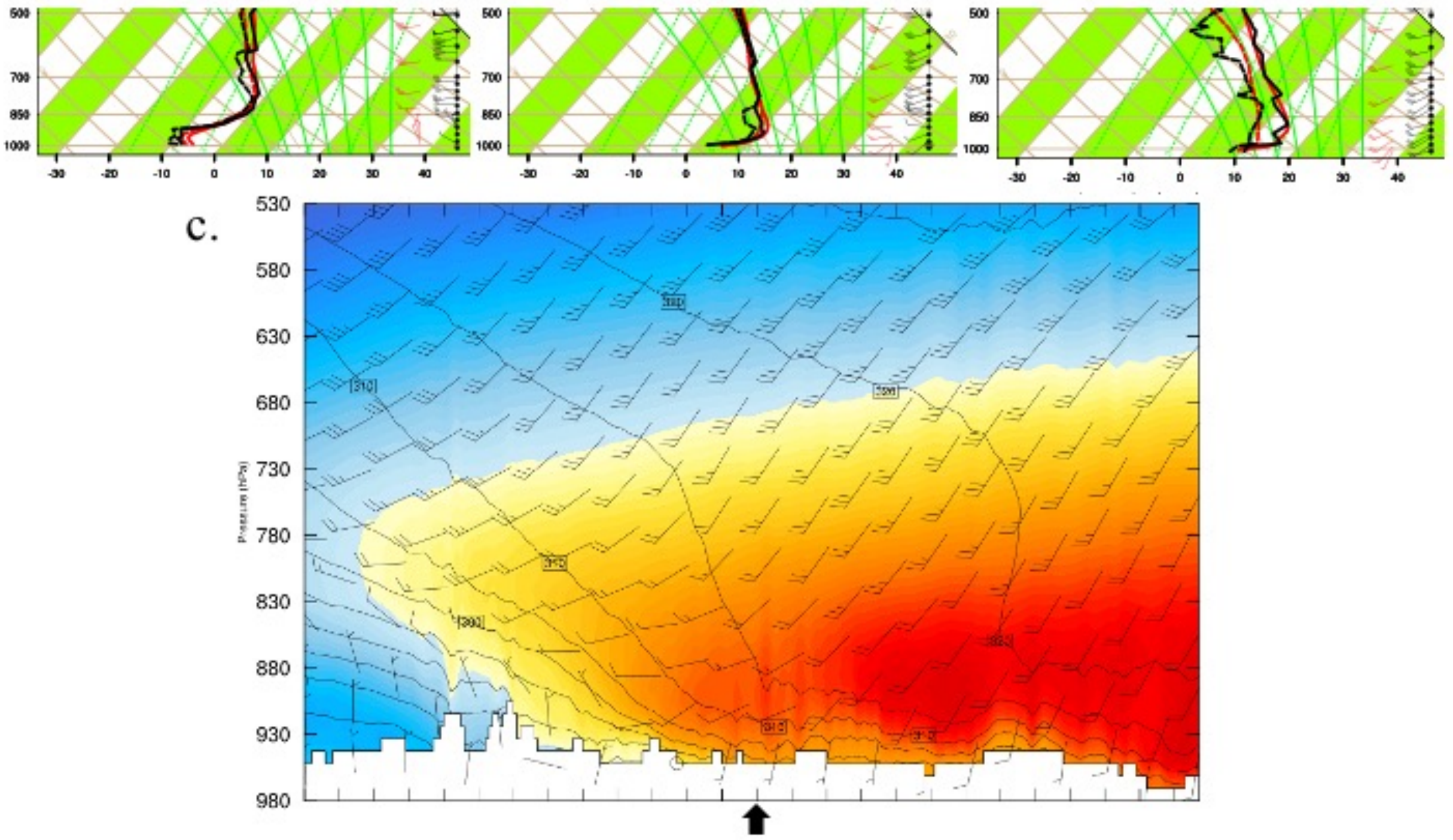

Fig. 7 cont.

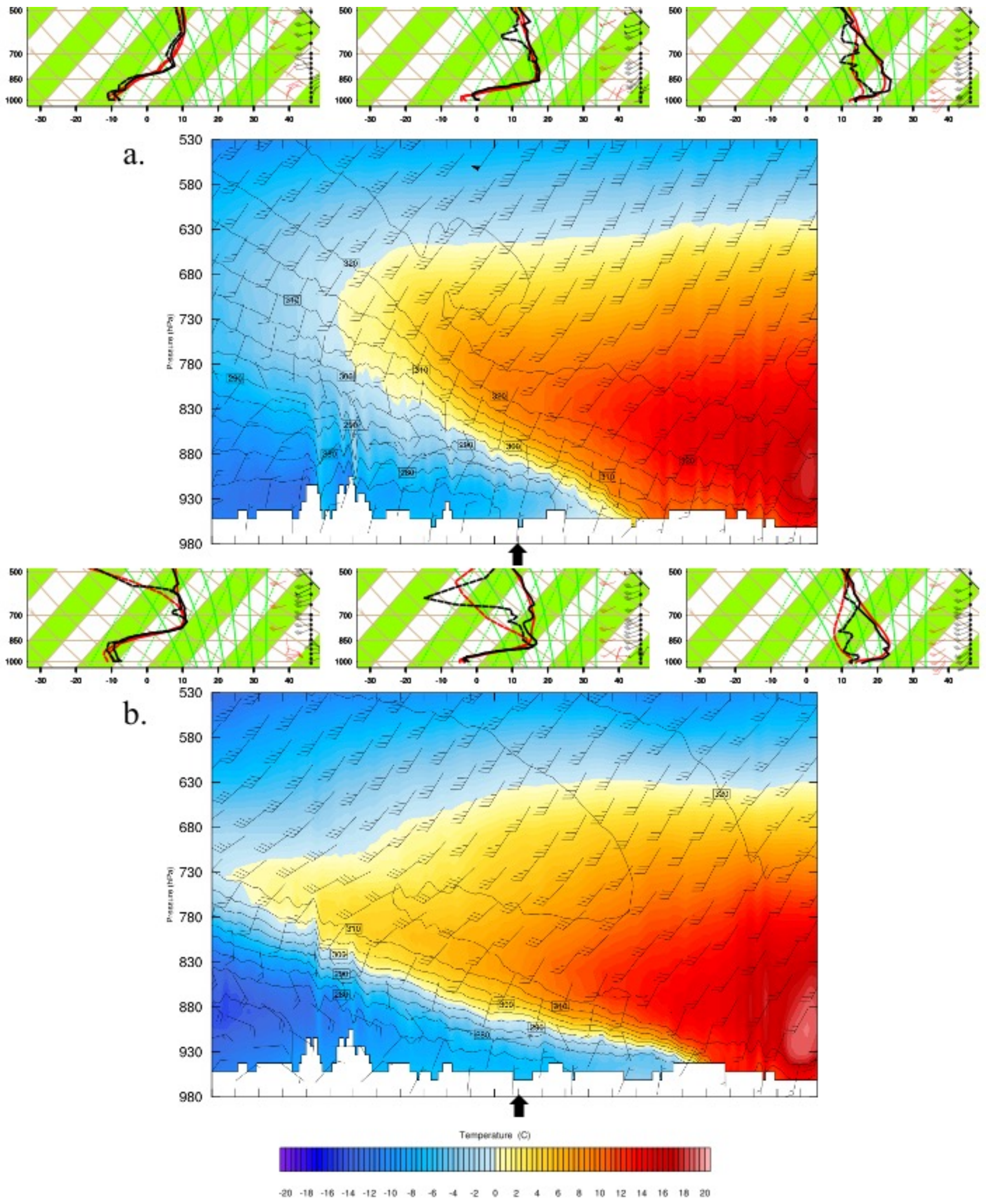

Fig. 8. As in Fig. 7, except cross sections and soundings at (a) 1200 UTC 22 Dec, (b) 0000 UTC and (c) 1200 UTC 23 Dec for the MYJ PBL simulation.

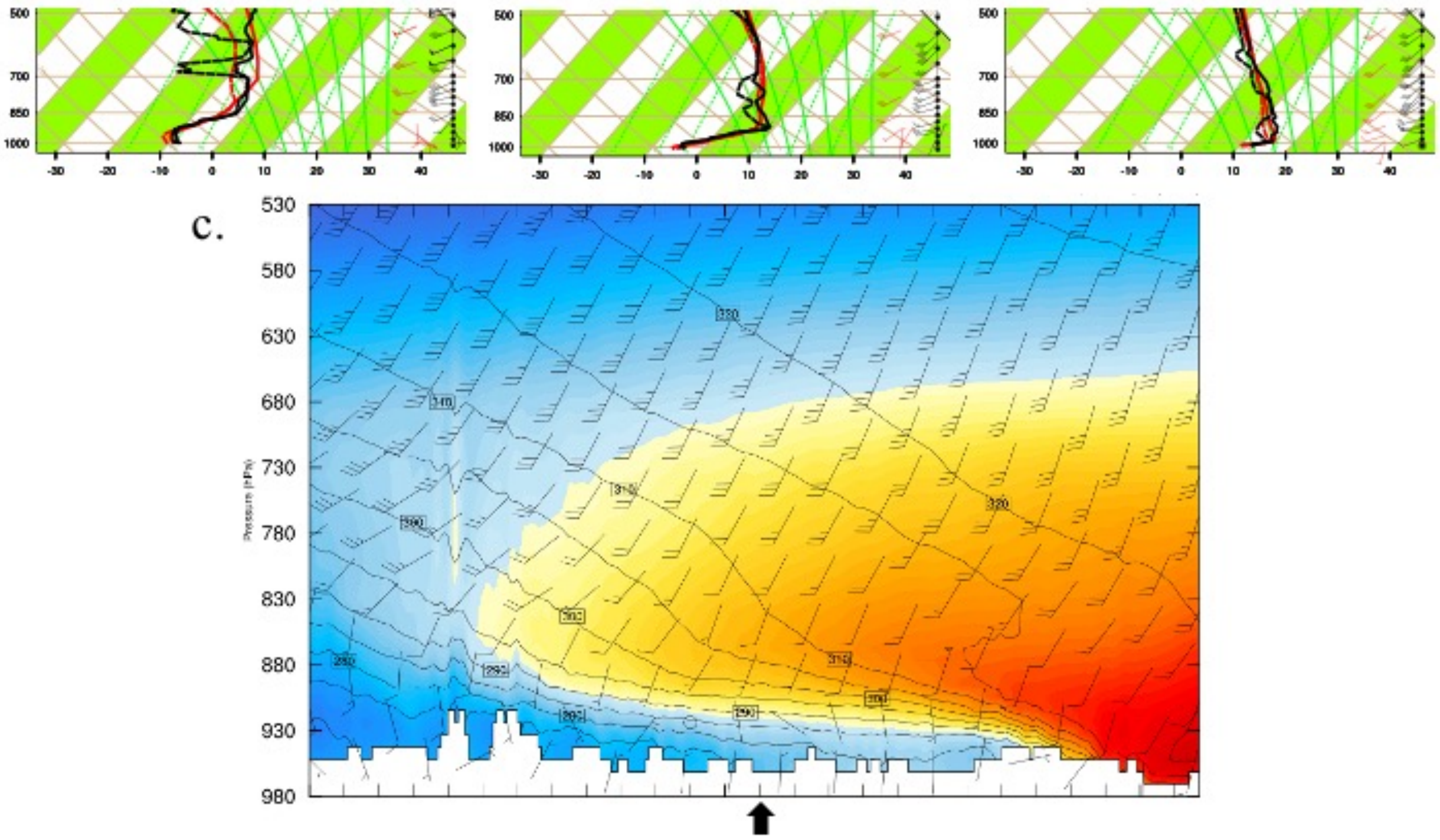

Fig. 8 cont.

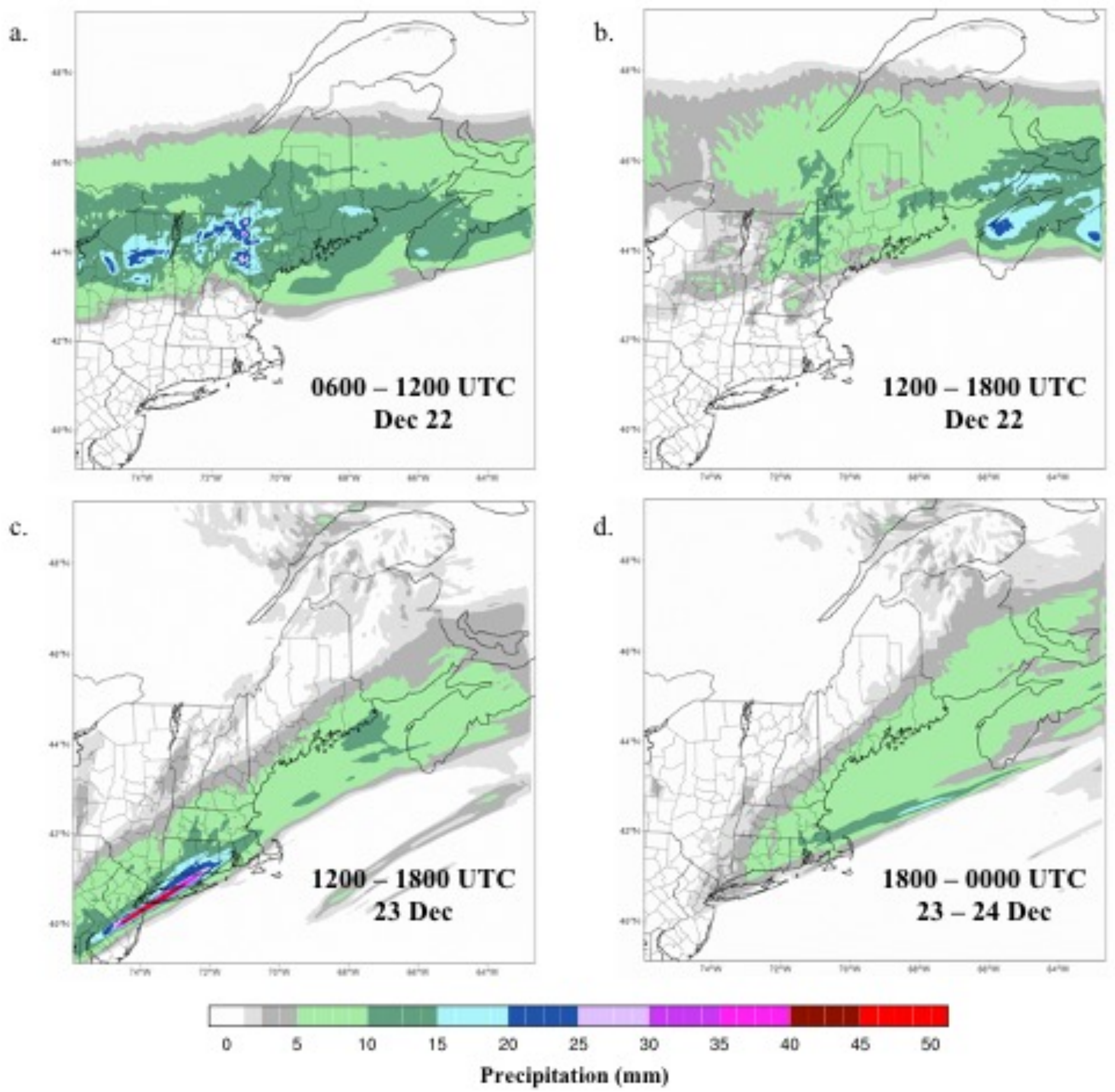

Fig. 9. Six-hour accumulated precipitation (mm) from (a) 0600 UTC to 1200 UTC, (b) 1200 to 1800 UTC 22 Dec, (c) 1200 UTC to 1800 UTC 23 Dec, and (d) 1800 UTC 23 Dec to 0000 UTC 24 Dec for the MYJ PBL simulation.

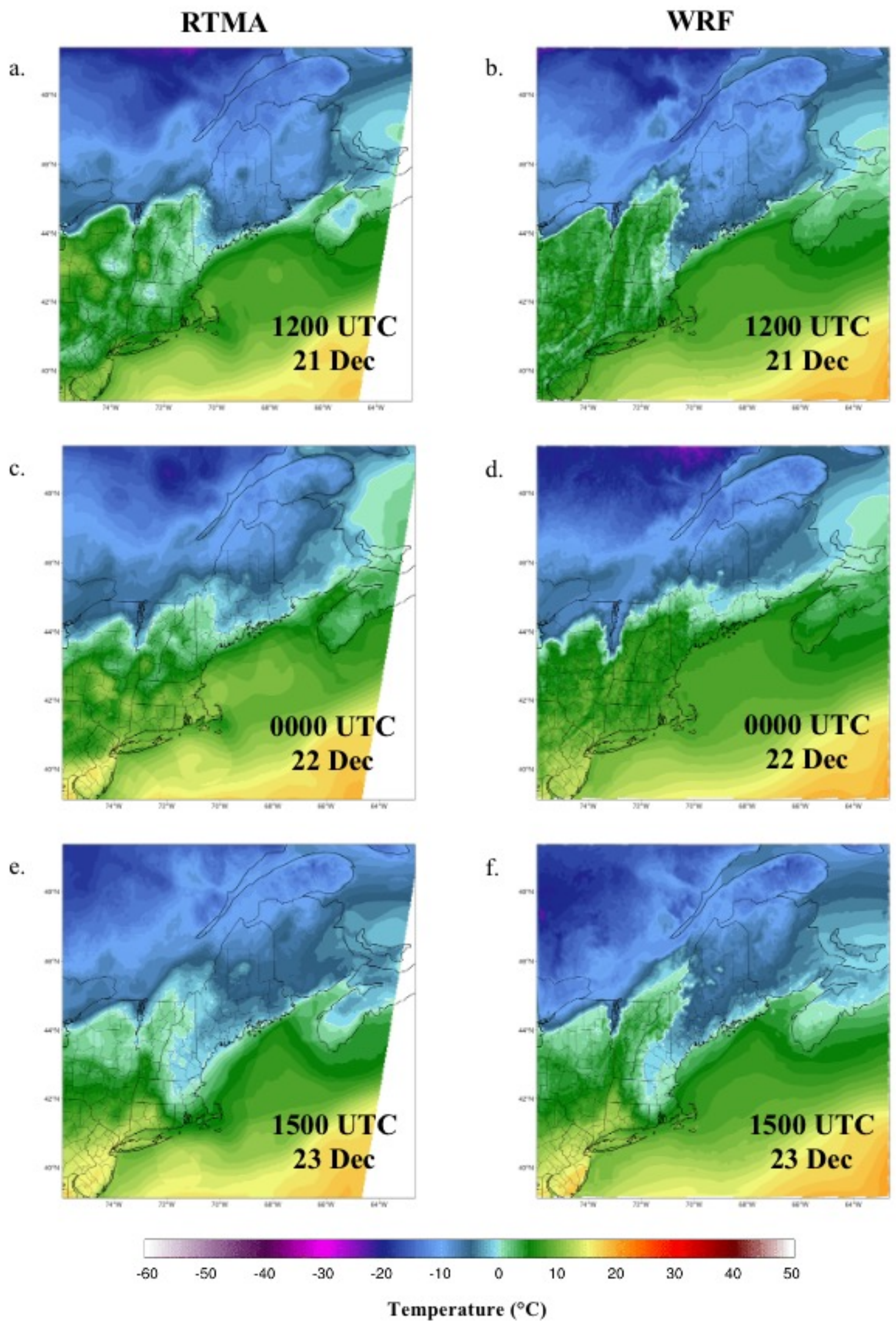

Fig. 10. Comparison of 2-meter temperature (°C) for RTMA (left) and MYJ PBL simulation (right) at 1200 UTC 21 Dec (a, b), 0000 UTC 22 Dec (c, d), and 1500 UTC 23 Dec (e, f).

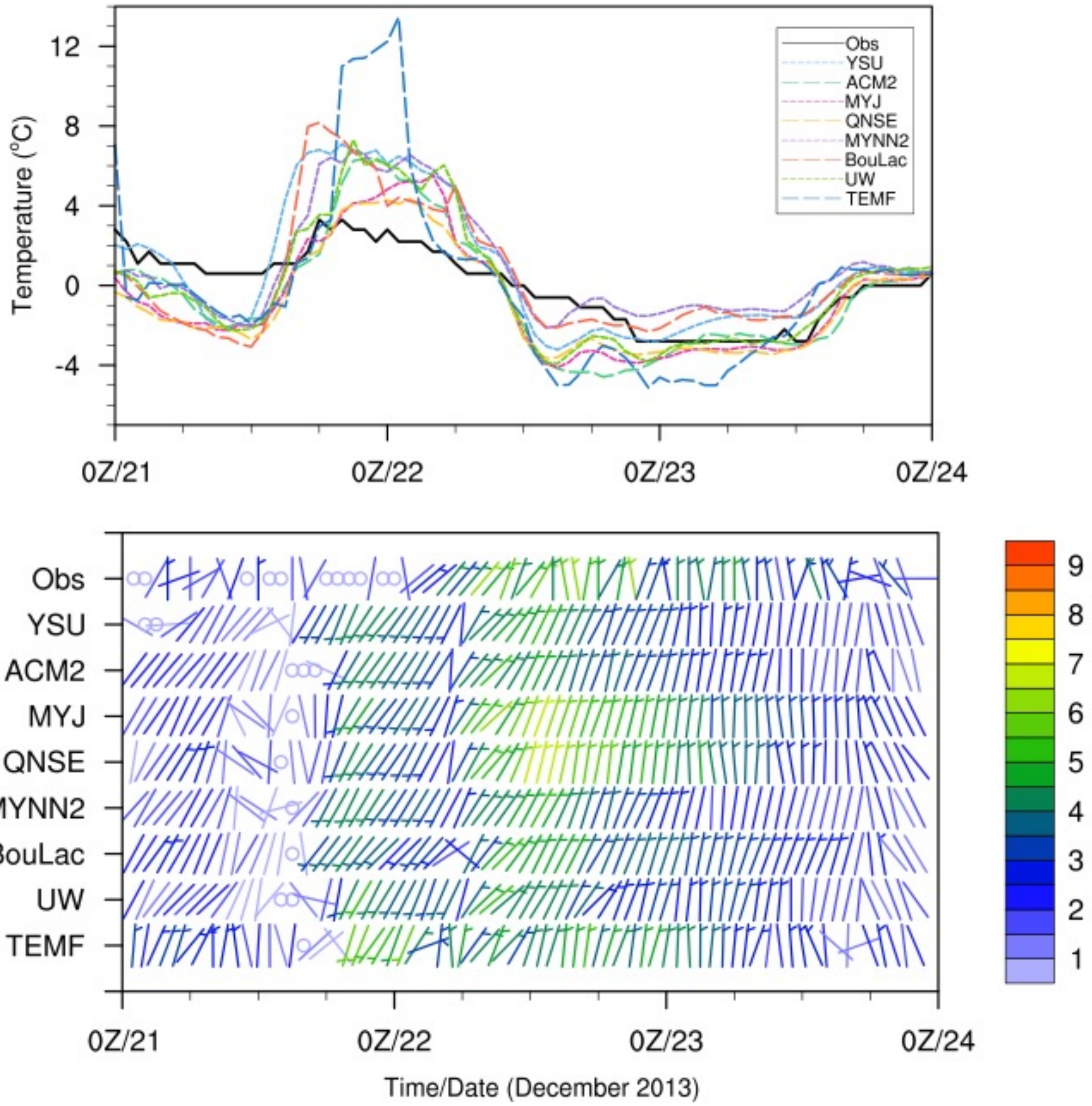

Fig. 11. Comparison of observed and modeled (PBL simulations) 2-meter temperature (top) and 10-meter wind (bottom, in $ms^{-1}$) time series for Portland, ME.

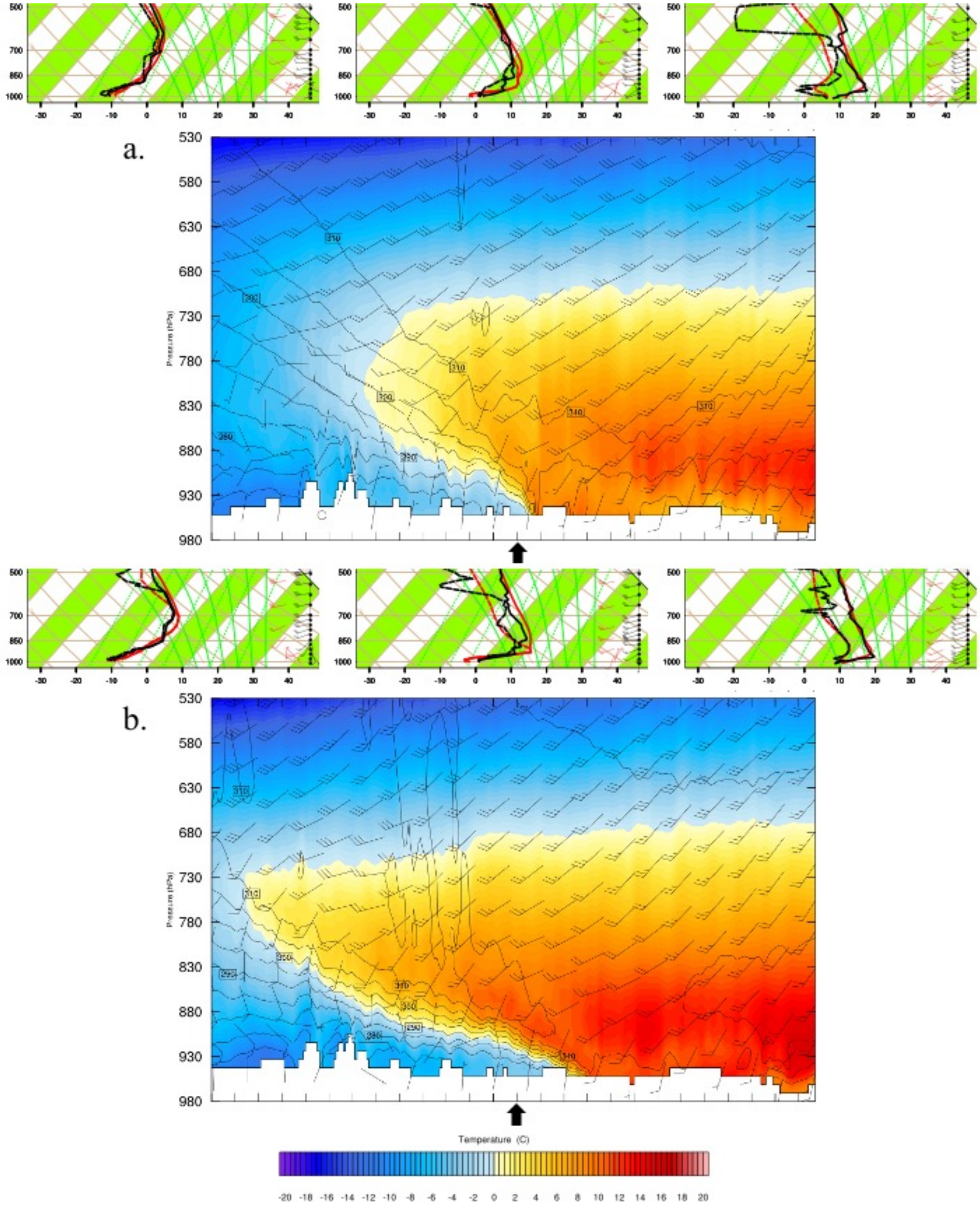

Fig. 12. Cross sections and soundings at (a) 0000 UTC and (b) 1200 UTC 21 Dec, and (c) 0000 UTC 22 Dec for the TEMF simulation.

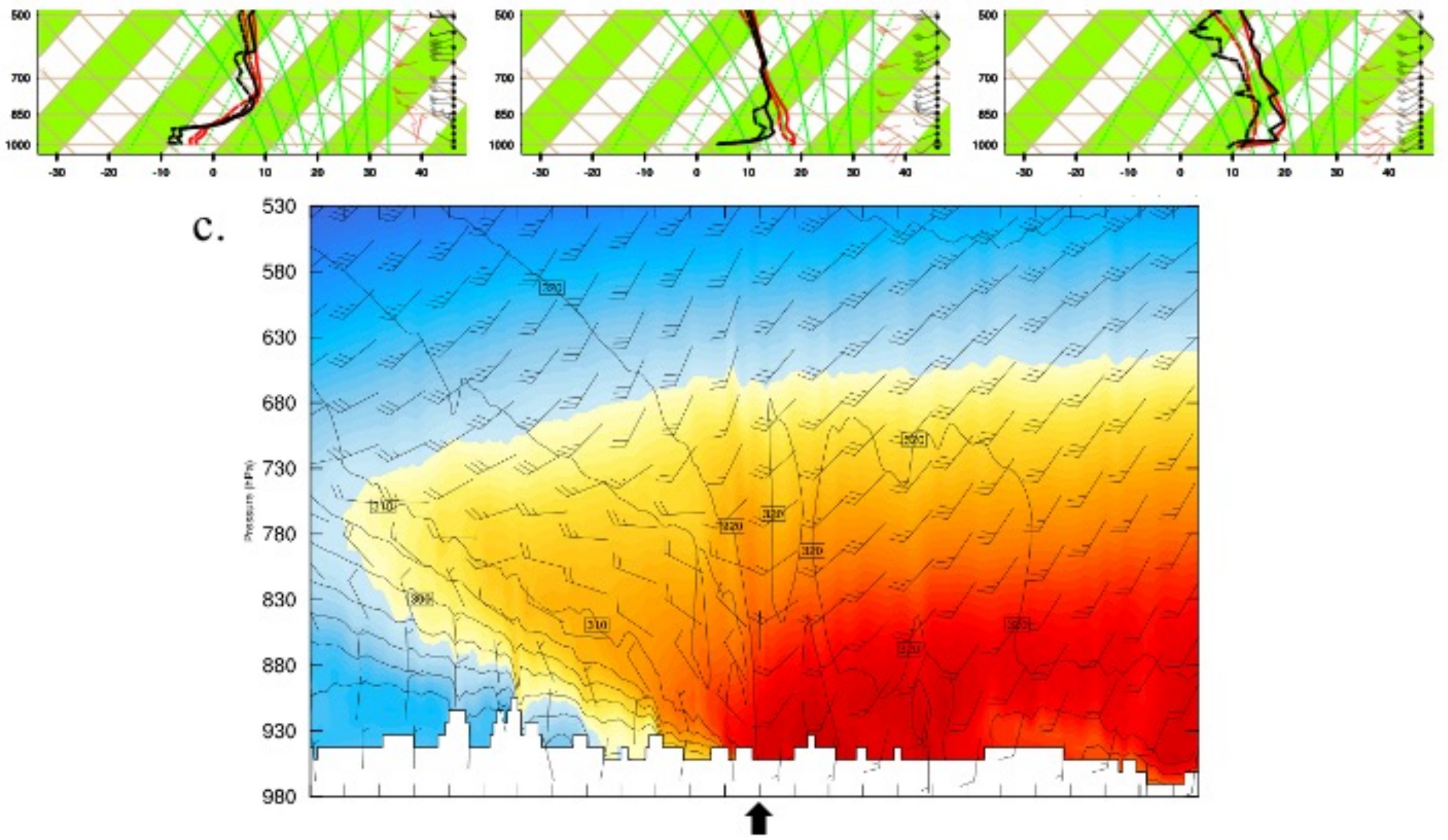

Fig. 12 cont.

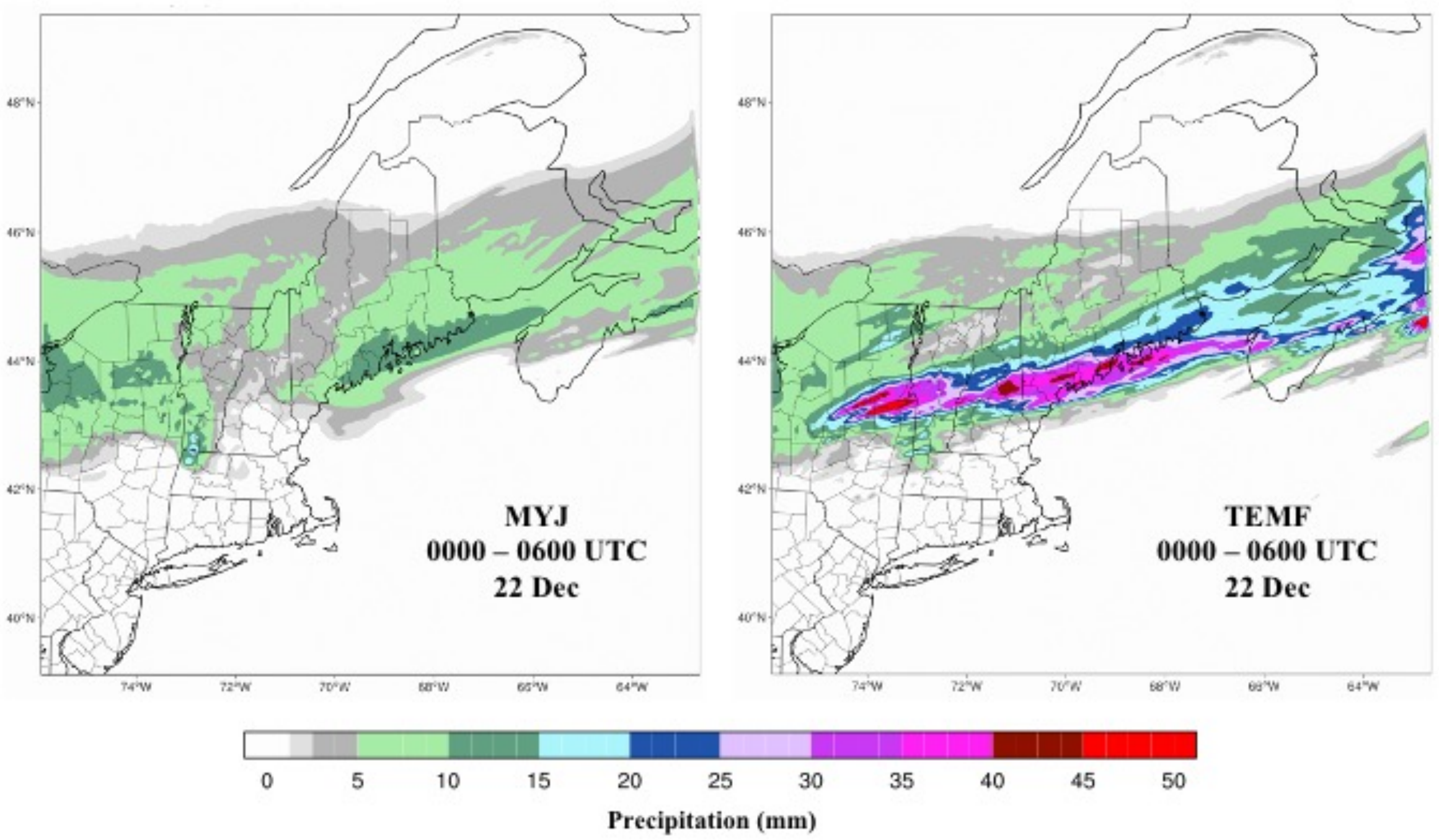

Fig. 13. Six-hour accumulated precipitation (mm) from 0000 UTC to 0600 UTC 22 Dec for the MYJ (left) and TEMF (right) PBL simulations.

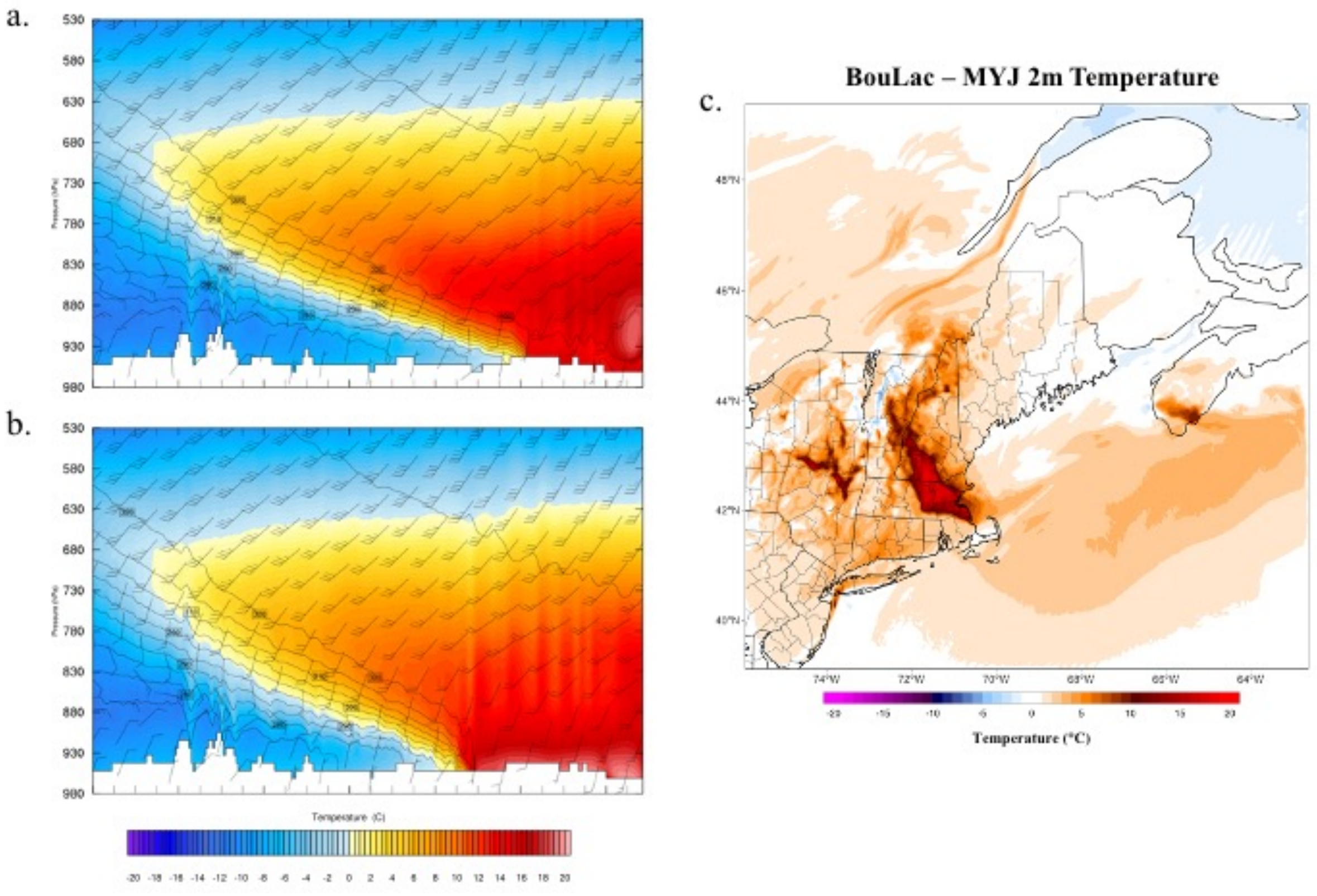

Fig. 14. Comparison of frontal passage variation between simulations: cross section of the (a) MYJ and (b) BouLac simulations, and (c) 2-meter temperature difference (°C) map between the BouLac and MYJ simulations at 1800 UTC 22 Dec.

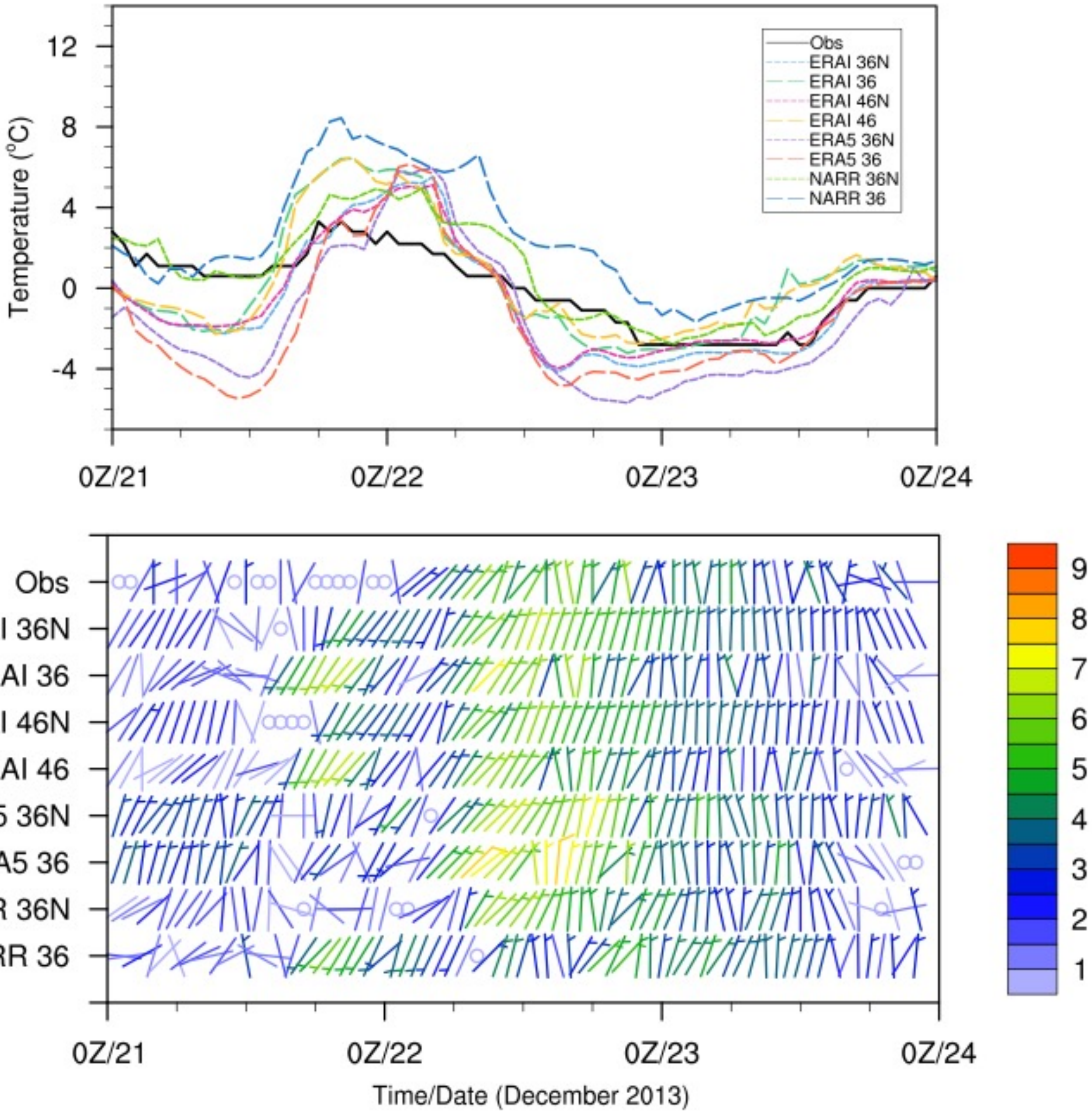

Fig. 15. As in Fig. 11, except for the model setup simulations. The ERAI 36N simulation corresponds to the MYJ PBL simulation.

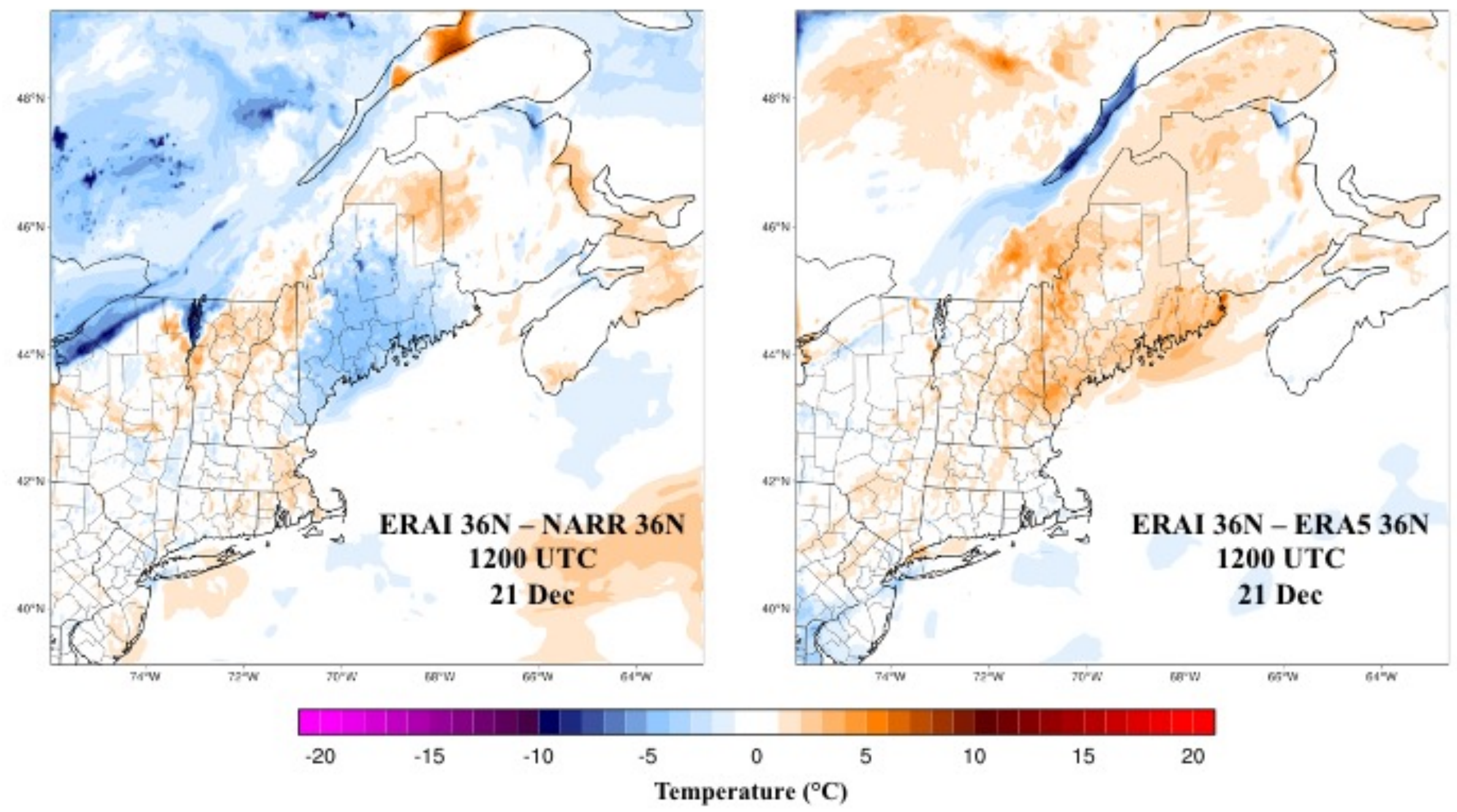

Fig. 16. Difference in WRF 2-meter temperatures (°C) between ERAI 36N and NARR 36N (left) and ERAI 36N and ERA5 36N (right) simulations at 1200 UTC 21 Dec 2013.

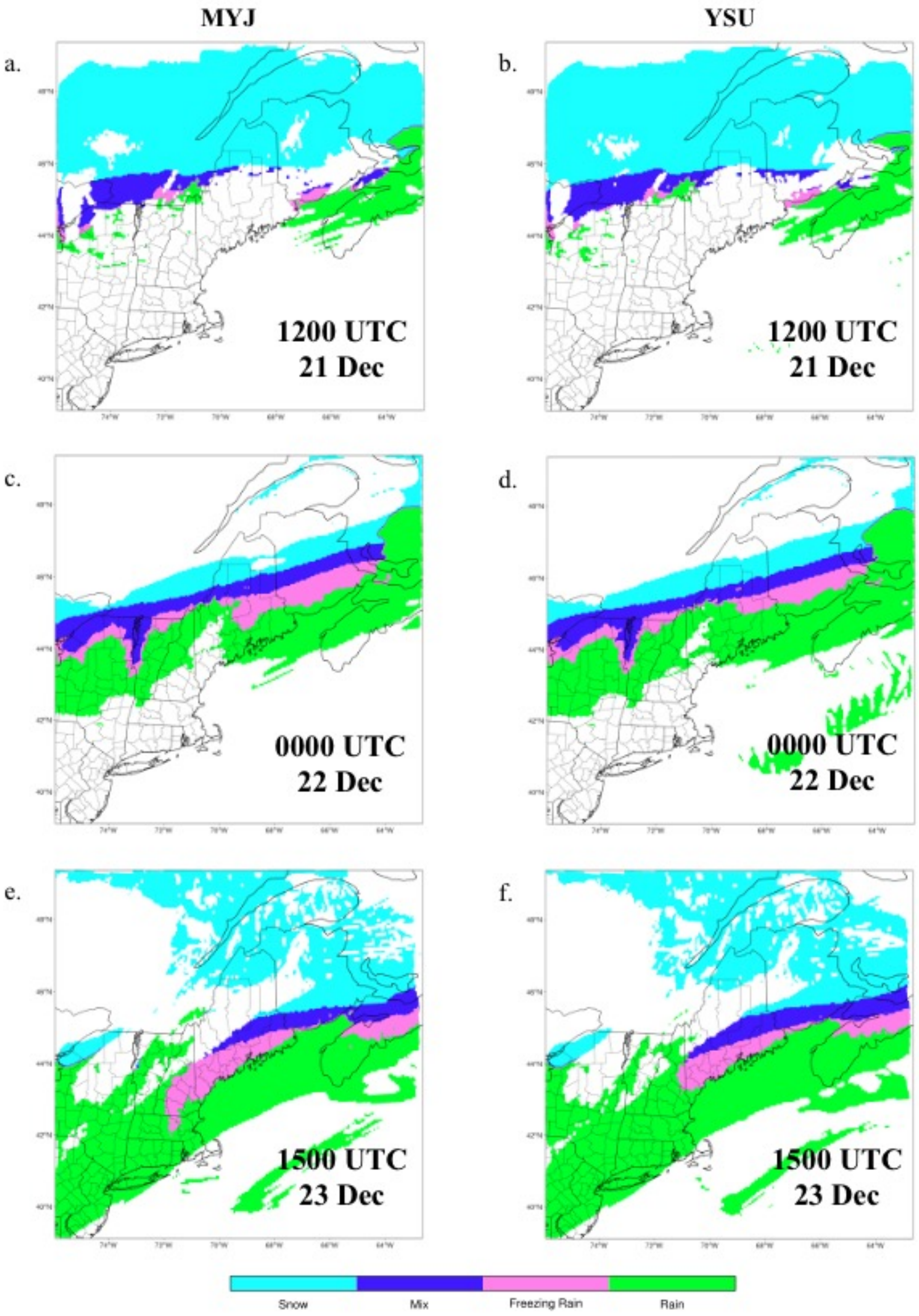

Fig. 17. Comparison of precipitation type for the MYJ (left) and YSU (right) simulations at 1200 UTC 21 Dec (a, b), 0000 UTC 22 Dec (c, d), and 1500 UTC 23 Dec (e, f).

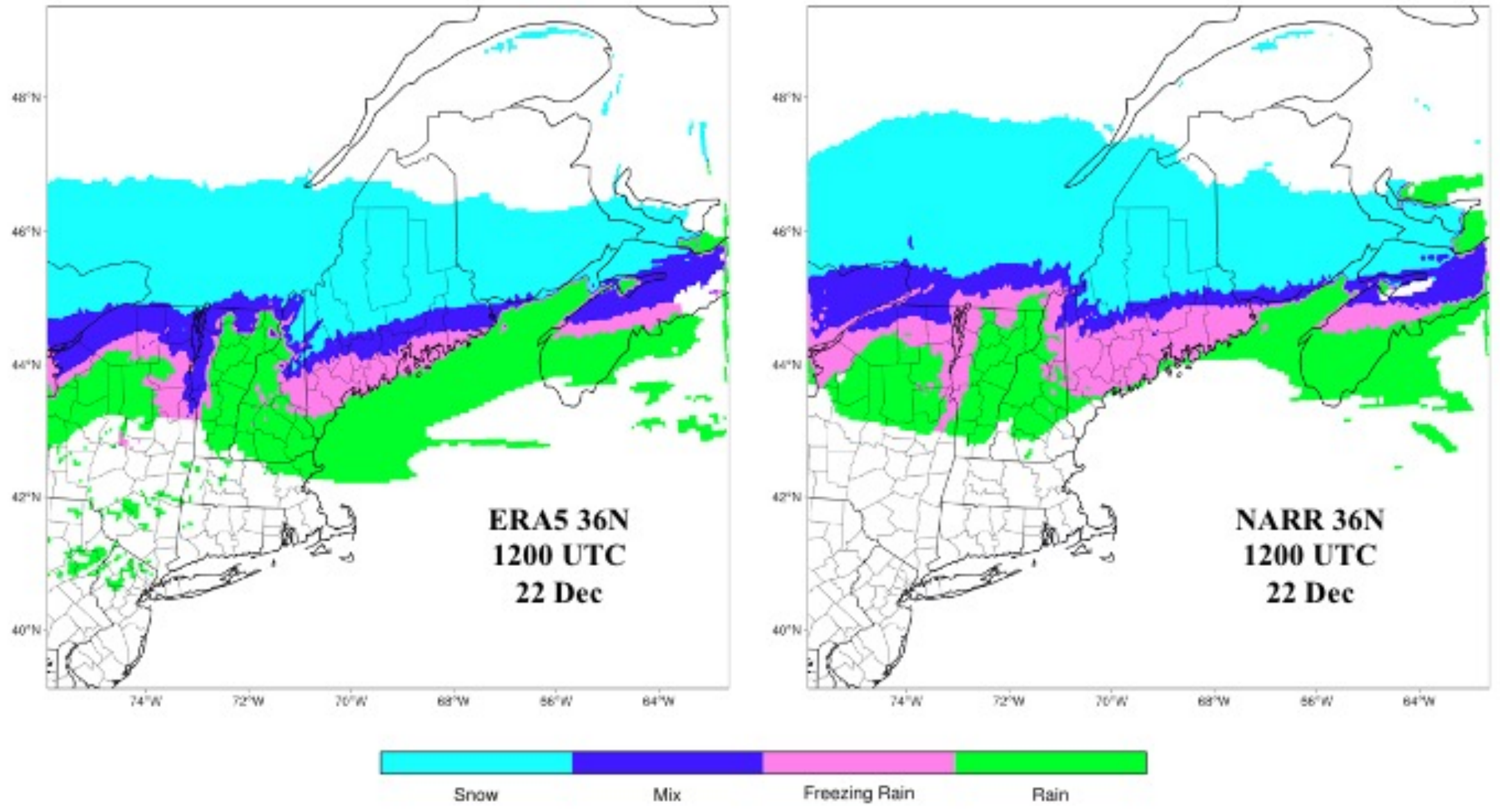

Fig. 18. Precipitation type at 1200 UTC 22 Dec for the ERA5 36N (left) and NARR 36N (right) simulations.